\pdfoutput=1

\documentclass[
  aps,
  prd,
  10pt,
  twocolumn,
  superscriptaddress,
  preprintnumbers,
  nofootinbib
]{revtex4-2}

\newif\ifdraft
\drafttrue

\usepackage[T1]{fontenc}

\usepackage{
  amsmath,
  amssymb,
  amsfonts,
  amsthm,
  mathtools,
  mathrsfs,
  bm,
  stmaryrd,
  dsfont,
  leftidx,
  accents,
  slashed,
  cancel,
  empheq
}

\usepackage{graphicx}
\graphicspath{{Images/}}

\usepackage{tikz}
\usepackage{tikz-3dplot}
\usepackage{pgfplots}

\pgfplotsset{compat=1.18}

\usetikzlibrary{
  arrows.meta,
  backgrounds,
  calc,
  decorations.markings,
  decorations.pathmorphing,
  decorations.pathreplacing,
  intersections,
  patterns,
  patterns.meta,
  shapes
}

\usepgfplotslibrary{fillbetween}

\tikzset{
  snakeit/.style={
    decorate,
    decoration=snake
  },
  zigzag/.style={
    decorate,
    decoration=zigzag
  }
}

\usepackage{array}
\usepackage{multirow}
\usepackage{tabularx}
\usepackage{diagbox}

\usepackage{subfig}

\usepackage{comment}
\usepackage{ifthen}
\usepackage{import}
\usepackage{wrapfig}
\usepackage[percent]{overpic}
\usepackage[most]{tcolorbox}
\usepackage{mdframed}
\usepackage{pifont}

\usepackage[dvipsnames]{xcolor}

\definecolor{rosy}{RGB}{230,235,252}
\definecolor{myframetitle}{RGB}{90,89,170}
\definecolor{myblocktitle}{RGB}{140,185,249}
\definecolor{mytitle}{RGB}{10,80,26}

\definecolor{darkgreen}{RGB}{27,130,45}
\definecolor{darkblue}{rgb}{0,0,0.3}
\definecolor{darkred}{rgb}{0.7,0,0}

\definecolor{lightgray}{RGB}{220,220,220}
\definecolor{darkpurple}{RGB}{108,0,217}

\definecolor{pink}{RGB}{190,20,100}
\definecolor{orang}{RGB}{193,63,0}
\definecolor{green}{RGB}{11,98,17}
\definecolor{darkpink}{RGB}{153,0,76}
\definecolor{bluegreen}{RGB}{0,102,102}
\definecolor{greenlagan}{RGB}{0,102,0}
\definecolor{redgreen}{RGB}{102,102,0}
\definecolor{Redgreen}{RGB}{153,76,0}

\definecolor{vividviolet}{rgb}{0.62,0.0,1.0}
\definecolor{amaranth}{rgb}{0.9,0.17,0.31}
\definecolor{palatinateblue}{rgb}{0.15,0.23,0.89}
\definecolor{brightpink}{rgb}{1.0,0.0,0.5}
\definecolor{cornflowerblue}{rgb}{0.39,0.58,0.93}
\definecolor{deepcarminepink}{rgb}{0.94,0.19,0.22}
\definecolor{radicalred}{rgb}{1.0,0.21,0.37}

\usepackage[
  colorlinks=true,
  linktoc=all,
  linkcolor=palatinateblue,
  citecolor=brightpink,
  urlcolor=amaranth
]{hyperref}

\ifdraft

\newcommand\hnote[1]{%
  \textcolor{blue}{\bfseries [HA:\,#1]}%
}

\newcommand\anote[1]{%
  \textcolor{magenta}{\bfseries [AL:\,#1]}%
}

\newenvironment{redinsert}
  {\begingroup\color{red}}
  {\endgroup}

\else

\newcommand\hnote[1]{}
\newcommand\anote[1]{}

\hypersetup{hidelinks}

\fi

\newcommand{\dif}{\mathop{}\!\mathrm{d}}

\tcbset{
  highlight math style={
    left=2mm,
    right=2mm,
    top=2mm,
    bottom=2mm
  }
}

\newcommand{\be}{\begin{equation}}
\newcommand{\ee}{\end{equation}}

\begin{document}

\title{On Integrable Structures on Non--compact Boundaries in Three--Dimensional Gravity
}


\author{H.~Adami}
\email{hadami@simis.cn}

\affiliation{
Center for Mathematics and Interdisciplinary Sciences,
Fudan University,
Shanghai 200433,
China
}

\affiliation{
Shanghai Institute for Mathematics and Interdisciplinary Sciences (SIMIS),
Shanghai 200433,
China
}

\author{,~K. ~Lara}
\email{klara@cecs.cl}

\affiliation{Centro de Estudios Cient\'{\i}ficos (CECs), Av. Arturo Prat 514, Valdivia, Chile.}
\affiliation{Facultad de Ingenier\'ia, Universidad San Sebastián, sede Valdivia, General Lagos 1163, Valdivia 5110693, Chile}

\author{A.~Latifi}
\email{latifi@qut.ac.ir}

\affiliation{
Department of Mechanics,
Qom University of Technology, Qom,
Iran
}

\author{R.~Meyer}
\email{rene.meyer@uni-wuerzburg.de}

\affiliation{Institute for Theoretical Physics and Astrophysics and W\"urzburg-Dresden Cluster of Excellence ctd.qmat,
Julius-Maximilians-Universit\"at W\"urzburg, 97074 W\"urzburg, Germany}

\affiliation{
	Shanghai Institute for Mathematics and Interdisciplinary Sciences (SIMIS),
	Shanghai 200433,
	China
}


\begin{abstract}
We study three-dimensional Einstein gravity with negative cosmological constant on non-compact spatial boundaries within the Chern--Simons formulation. Using an exact fluid/gravity correspondence, we derive a closed radial flow equation for the quasi-local stress tensor and show that it realizes the holographic {$\sqrt{T\bar T}$} deformation at finite cutoff. We further develop the inverse-scattering description of the boundary dynamics, identifying the gravitational interpretation of the associated spectral data and analyzing the finite-cutoff deformation of soliton solutions. Although the boundary evolution is governed by an integrable bi-Hamiltonian hierarchy, we show that the radial flow itself is not Hamiltonian with respect to the canonical Poisson structure. Our results establish a unified framework connecting integrability, quasi-local gravitational observables, inverse scattering, and finite-cutoff holography on non-compact boundaries.
\end{abstract}

\maketitle

\tableofcontents

\section{Introduction}

Three-dimensional Einstein gravity provides one of the simplest yet
most instructive laboratories for exploring both classical and quantum
aspects of gravitation. Although the theory possesses no local
propagating gravitational degrees of freedom, its solution space is far
from trivial. Every vacuum solution is locally a maximally symmetric
spacetime, with the cosmological constant determining whether the local
geometry is Minkowski, de Sitter, or anti-de Sitter. Nevertheless,
global identifications, nontrivial topology, holonomies, and the
presence of boundaries generate a rich space of physically inequivalent
configurations. In particular, asymptotically AdS$_3$ spacetimes admit
an infinite-dimensional asymptotic symmetry algebra whose canonical
charges encode the dynamical information of the theory. This remarkable
interplay between topology, boundary conditions, and symmetry has made
three-dimensional gravity a cornerstone in the study of holography,
black-hole thermodynamics, and exactly solvable models of quantum
gravity
~\cite{Brown:1986nw,Henneaux:1985tv,Witten:1988hc,Carlip:1998uc,Carlip:2005zn,Banados:1992wn,Banados:1993ur}.

A particularly elegant formulation of three-dimensional gravity with
negative cosmological constant is provided by the
Chern--Simons description, where the Einstein--Hilbert action can be
rewritten as the difference of two Chern--Simons actions with gauge
group
$\mathrm{SL}(2,\mathbb{R})\times\mathrm{SL}(2,\mathbb{R})$
~\cite{Achucarro:1987vz,Witten:1988hc}. In this language the field
equations simply require the gauge connections to be flat, rendering
the local dynamics topological. Consequently, the physical degrees of
freedom are entirely associated with global properties of the gauge
bundle and, crucially, with the boundary conditions imposed on the
connection. The Chern--Simons formulation therefore provides a natural
framework for classifying admissible phase spaces, identifying
asymptotic symmetries, constructing conserved charges, and revealing
the Hamiltonian structure underlying the theory
~\cite{Elitzur:1989nr,Coussaert:1995zp,Banados:1994tn,Banados:1998gg,Compere:2018aar}.

The importance of boundary conditions extends well beyond the
construction of asymptotic symmetry algebras. Different admissible
boundary conditions define distinct gravitational phase spaces,
characterized by different sets of boundary degrees of freedom,
canonical generators, and dynamical equations. Over the past decade it
has become increasingly clear that many of these boundary dynamics are
governed by integrable nonlinear evolution equations. Rather than
representing isolated examples, these integrable structures appear to
constitute a generic feature of broad classes of boundary conditions in
three-dimensional gravity, establishing an unexpected bridge between
classical gravitation and the mathematical theory of integrable systems.

A systematic realization of this correspondence was developed in
~\cite{Perez:2016vqo}, where a family of boundary conditions was
constructed by identifying the chemical potentials conjugate to the
boundary currents with the Gel'fand--Dikii polynomials of the Korteweg--de Vries (KdV)
hierarchy. Within this framework, the Einstein equations projected onto
the boundary reduce exactly to two independent chiral copies of the
$k$-th flow of the KdV hierarchy, while the canonical
gravitational charges coincide with the commuting Hamiltonians of the
integrable system. The infinite sequence of commuting conserved charges,
the bi-Hamiltonian structure, and the associated recursion relations
thus emerge naturally from the gravitational phase space itself,
providing a striking realization of integrability directly within the
dynamics of Einstein gravity. This correspondence further establishes a
precise map between geometric data specifying the asymptotic solution
space and the hierarchy of commuting Hamiltonian flows that characterize
the KdV system.

Subsequent developments have considerably broadened the scope of this
correspondence. Allowing the chemical potentials to depend locally on
the dynamical fields, instead of restricting them to the polynomial
Gel'fand--Dikii family, gives rise to boundary dynamics governed by the
Gardner hierarchy, thereby interpolating continuously between the KdV
and modified KdV equations~\cite{Ojeda:2019xih}. Closely related
constructions have been extended to higher-spin theories, where the
underlying Drinfeld--Sokolov reduction naturally leads to generalized
integrable hierarchies, including the modified Boussinesq system
~\cite{Ojeda:2020bgz}. More generally, the Drinfeld--Sokolov framework
provides a unified algebraic perspective on the relation between
boundary conditions, chemical potentials, and the resulting integrable
equations, clarifying how different gravitational phase spaces are
organized according to the associated Hamiltonian hierarchies
~\cite{Gonzalez:2018jgp,Campoleoni:2010zq,Campoleoni:2011hg}.

The correspondence between three-dimensional gravity and integrable
systems extends significantly beyond the KdV and Gardner hierarchies.
A particularly rich example arises from boundary conditions whose
chemical potentials generate the Ablowitz--Kaup--Newell--Segur (AKNS)
hierarchy. In this case, the boundary dynamics are governed by the
AKNS equations, while distinct gravitational configurations, including
conical defects and extremal black holes, are naturally distinguished
by the conjugacy class of the corresponding holonomies
~\cite{Cardenas:2021vwo,Cardenas:2025qqi}. This observation further illustrates that the
choice of boundary conditions not only determines the asymptotic
symmetry algebra but also selects the underlying integrable structure,
thereby providing a direct geometric realization of different
integrable hierarchies within the gravitational phase space.

The scope of these constructions has continued to expand in several
directions. Perturbative deformations of the Brown--Henneaux boundary
conditions incorporating finite-$1/c$ corrections have recently been
shown to give rise to the Dym hierarchy, revealing that integrability
persists even beyond the conventional semiclassical regime
~\cite{Lara:2024cie,Pino:2025crn}. Related developments have also clarified the
interplay between integrable boundary dynamics, higher-spin gauge
theories, and non-relativistic holography, where Lifshitz boundary
conditions lead to generalized hierarchies possessing anisotropic
scaling symmetry
~\cite{Compere:2013gja,Melnikov:2018fhb,Ojeda:2020bgz,Arenas-Henriquez:2024ypo}. Taken
together, these results suggest that integrability is not tied to a
particular choice of boundary conditions but rather represents a
generic organizing principle governing large classes of admissible
gravitational phase spaces.

Analogous phenomena also arise outside the AdS$_3$ setting. In the
limit of vanishing cosmological constant, suitable boundary conditions
lead to integrable systems whose Poisson structure is governed by the
$\mathrm{BMS}_3$ algebra rather than two copies of the Virasoro
algebra~\cite{Fuentealba:2017omf}. The resulting hierarchy provides a
natural integrable realization of asymptotically flat gravity and
demonstrates that the emergence of nonlinear integrable dynamics is
not restricted to conformal boundaries. Instead, it appears to reflect
a deeper relationship between asymptotic symmetry algebras,
Hamiltonian structures, and admissible gravitational boundary
conditions.

A closely related perspective emerges from the study of boundary
conditions imposed near black-hole horizons. While soft-hair boundary
conditions possess an extensive degeneracy associated with
near-horizon symmetry transformations, KdV-type boundary conditions
lift this degeneracy by promoting the boundary fields to evolve
according to an integrable hierarchy
~\cite{Afshar:2016wfy,Grumiller:2019tyl}. The resulting black-hole
solutions carry an infinite family of commuting charges inherited from
the corresponding integrable system, and their thermodynamic
properties have been investigated in detail
~\cite{Erices:2019onl,Dymarsky:2020tjh}. These developments reinforce
the view that integrable structures provide a unifying framework for
understanding both asymptotic and near-horizon gravitational dynamics.

Although many investigations have focused on compact spatial slices,
non-compact boundaries introduce qualitatively new mathematical
structures. In this setting the inverse scattering transform becomes
available, allowing the nonlinear evolution equations governing the
boundary dynamics to be solved through spectral methods. In
~\cite{Adami:2025pfk}, the KdV boundary conditions were extended to
non-compact spatial slices, where the associated gravitational phase
space naturally decomposes into solitonic and radiative sectors. The
reflectionless sector is described by a forced KdV hierarchy that
admits an exact solution through the
Gelfand--Levitan--Marchenko reconstruction, while the continuous
spectrum gives rise to dispersive radiation exhibiting universal
large-time asymptotics. This construction provides the first explicit
realization of inverse scattering techniques directly within the
boundary dynamics of three-dimensional Einstein gravity and endows the
gravitational phase space with a transparent spectral
interpretation.

These developments collectively reveal a remarkably rich interplay
between gravitational dynamics, infinite-dimensional symmetry
algebras, and classical integrability. Different choices of boundary
conditions generate different Hamiltonian hierarchies, while the
underlying Chern--Simons formulation provides a common geometric
framework in which these apparently distinct systems arise. The
boundary currents simultaneously encode the boundary gravitational
data and the dynamical variables of the associated integrable model,
thereby establishing a precise correspondence between gravitational
phase spaces and nonlinear evolution equations. From this viewpoint,
three-dimensional gravity furnishes a natural arena in which methods
from integrable systems—including bi-Hamiltonian geometry, Lax pairs,
recursion operators, and inverse scattering—acquire direct
gravitational significance. Conversely, gravitational considerations
offer a geometric interpretation of the conserved quantities, spectral
data, and Hamiltonian structures characteristic of integrable
hierarchies. This dual perspective motivates the search for new
connections between holography, boundary dynamics, and integrability,
particularly in situations where additional physical ingredients, such
as finite radial cutoffs or deformed holographic dualities, are
present.

A conceptually independent, yet ultimately complementary, line of
research has emerged from the study of holography at a finite radial
cutoff. In the conventional AdS/CFT correspondence the dual quantum
field theory is defined at the conformal boundary, corresponding to the
ultraviolet fixed point of the renormalization group. Moving the
holographic screen to a finite radial position instead provides a
description of the dual theory at a finite energy scale, thereby
offering a geometric realization of Wilsonian renormalization group
flow in holography
~\cite{deBoer:1999xf,Bianchi:2001kw,Skenderis:2002wp,Papadimitriou:2004ap,Heemskerk:2010hk,Faulkner:2010jy,Adami:2025pqr}. This
perspective has attracted considerable attention because it enables the
study of irrelevant deformations, effective field theories away from
conformal fixed points, and the emergence of spacetime dynamics along
the holographic renormalization group trajectory.

A major advance in this direction was the proposal that introducing a
finite radial cutoff in asymptotically AdS$_3$ gravity is holographically
equivalent to deforming the dual two-dimensional conformal field theory
by the composite $T\bar T$ operator
~\cite{McGough:2016lol}. Independently, it was shown that the
$T\bar T$ operator defines a distinguished class of solvable
irrelevant deformations of two-dimensional quantum field theories,
generated by the determinant of the stress tensor
~\cite{Zamolodchikov:2004ce,Smirnov:2016lqw,Cavaglia:2016oda}. Despite
being irrelevant in the renormalization group sense, the deformation
preserves an exceptional degree of solvability: the finite-volume energy
spectrum satisfies an exact equation, many observables can
be determined non-perturbatively, and several integrable properties of
the undeformed theory survive along the flow. These remarkable features
have established the $T\bar T$ deformation as one of the best
understood examples of an exactly solvable irrelevant deformation in
quantum field theory.

The finite-cutoff interpretation has subsequently been investigated from
a variety of complementary viewpoints. Evidence for the holographic
dictionary has been obtained through the exact matching of deformed
energy spectra, thermodynamic quantities, transport coefficients,
correlation functions, and the radial evolution of the Brown--York
stress tensor
~\cite{Hartman:2018tkw,Aharony:2018bad,Taylor:2018xcy,Kraus:2018xrn,Guica:2019nzm}. Generalizations to theories with
additional conserved currents, higher-spin fields, supersymmetry,
non-relativistic symmetries, and related irrelevant deformations have
further broadened the scope of this framework, highlighting the central
role played by solvable deformations in modern holography
~\cite{Guica:2017lia,Chakraborty:2018vja,Apolo:2018qpq}. Collectively,
these developments provide compelling evidence that finite radial
cutoffs furnish a geometric realization of exactly solvable
renormalization group flows.

Most existing derivations of the holographic $T\bar T$ correspondence,
however, are formulated either in terms of the finite-volume energy
spectrum of the dual theory or through trace relations satisfied by the
renormalized boundary stress tensor
~\cite{McGough:2016lol,Hartman:2018tkw,Guica:2019nzm}. While these
approaches successfully characterize many physical consequences of the
deformation, they typically provide only indirect information about the
complete radial evolution of local observables. In particular, the
radial dependence of the quasi-local stress tensor is often reconstructed
perturbatively through holographic renormalization or near-boundary
expansions, making an exact non-perturbative description difficult to
obtain except in highly symmetric situations.

The fluid/gravity correspondence offers a complementary framework for
addressing this problem. In this approach the Brown--York stress tensor
defined on a timelike hypersurface at finite radial position is
interpreted as the energy-momentum tensor of an effective relativistic
fluid propagating on the induced geometry
~\cite{Bhattacharyya:2008xc,Bhattacharyya:2008ji,Erdmenger:2008rm,Banerjee:2008th,Rangamani:2009xk,Hubeny:2011hd}. The radial
Einstein equations are then mapped into evolution equations governing
the fluid variables, while radial translations acquire the
interpretation of holographic renormalization group flow
~\cite{deBoer:1999xf,Skenderis:2002wp,Papadimitriou:2004ap}. This
correspondence has proven to be a powerful tool for relating geometric
properties of gravitational solutions to hydrodynamic transport,
non-equilibrium dynamics, and effective field theories at finite energy
scales.

Three-dimensional gravity occupies a particularly distinguished position
within this framework. Owing to the absence of local bulk degrees of
freedom, the complete bulk geometry can be reconstructed exactly from
boundary data, allowing the quasi-local Brown--York stress tensor to be
determined in closed form throughout the bulk rather than only
asymptotically. Consequently, the fluid variables become explicit
algebraic functions of the radial coordinate and the underlying chiral
boundary currents
~\cite{Ojeda:2019xih,Grumiller:2019tyl}. This exceptional simplification
opens the possibility of deriving exact radial flow equations valid at
arbitrary cutoff, thereby providing a non-perturbative realization of
finite-cutoff holography that is largely inaccessible in higher
dimensions. It is precisely this feature that makes
three-dimensional gravity an ideal setting in which to investigate the
interplay between integrable boundary dynamics, quasi-local
gravitational observables, and solvable holographic renormalization
group flows.

Against this background, an important question naturally arises:
whether the integrable structures governing the boundary dynamics of
three-dimensional gravity admit a coherent description at finite radial
cutoff, where holographic renormalization group flow and
{$\sqrt{T\bar{T}}$}-deformed dynamics become essential. While integrability and
finite-cutoff holography have each been studied extensively, the
relationship between these two frameworks remains only partially
understood. In particular, it is not known to what extent the
Hamiltonian structures, spectral data, and inverse-scattering
description characterizing the boundary integrable hierarchy survive
under exact radial evolution, nor how the corresponding quasi-local
gravitational observables evolve along the holographic flow. Addressing
these questions is one objective of the present work.

In this paper we investigate the interplay between integrable boundary
dynamics and finite-cutoff holography within the Chern--Simons
formulation of three-dimensional Einstein gravity on non-compact
spatial slices. Starting from the diagonal reduction of the gauge
connections, we combine the exact fluid/gravity dictionary with the
Hamiltonian formulation of the boundary theory to obtain a
non-perturbative description of the radial evolution of quasi-local
gravitational observables. Unlike conventional approaches based on
near-boundary expansions, our construction is valid at arbitrary radial
position and therefore captures the complete finite-cutoff dynamics in
closed form.

A central result of this work is the derivation of an exact local
radial flow equation for the quasi-local energy density. We show that
this equation possesses precisely the structure expected for a {$\sqrt{T\bar{T}}$} deformation and therefore provides a direct gravitational
realization of the finite-cutoff flow entirely in terms of quasi-local
variables. Since the Brown--York stress tensor is known exactly in the
present framework, the resulting flow equation is non-perturbative and
resums all orders in the radial cutoff. This provides an explicit
realization of holographic renormalization group evolution that is
considerably stronger than the perturbative constructions usually
available in asymptotically AdS gravity.

The second objective of this paper is to incorporate the
inverse-scattering description of the boundary integrable hierarchy into
the finite-cutoff framework. Exploiting the non-compact nature of the
spatial boundary, we formulate the associated spectral problem and
analyze the evolution of the scattering data under radial flow. This
allows us to identify the gravitational interpretation of the spectral
data appearing in the inverse-scattering transform and to determine how
the soliton and radiative sectors of the theory are modified away from
the asymptotic boundary. In particular, we investigate the deformation
of reflectionless configurations and characterize the fate of the
multi-soliton sector at finite cutoff.

Further, we analyze the identification of a remarkable
distinction between the quasi-local momentum and energy densities. We
show that the momentum density satisfies an exact conservation law
under the full two-sector integrable dynamics, whereas the energy
density generically fails to be conserved. The origin of this difference
is traced to an explicitly computable coupling between the left- and
right-moving sectors that appears in the finite-cutoff theory but is
absent in the asymptotic description. This provides a precise mechanism
through which radial evolution modifies the conservation properties of
the boundary observables while preserving the underlying integrable
structure of the chiral dynamics.

We furthermore examine the mathematical structure of the radial flow
itself. Although the boundary time evolution is governed by an
integrable bi-Hamiltonian hierarchy, we demonstrate that the radial
evolution cannot be generated as a Hamiltonian flow with respect to the
same Poisson structure. {More precisely, we argue that no local Hamiltonian functional reproduces the exact
finite-cutoff flow within the canonical Poisson algebra of the boundary
currents.} Nevertheless, the radial evolution remains exactly solvable,
and we clarify the sense in which solvability and Hamiltonian
integrability become distinct notions once finite-cutoff effects are
taken into account.

Beyond their immediate application to three-dimensional gravity, our
results provide a unified framework connecting integrable hierarchies,
inverse scattering, quasi-local gravitational observables, and
finite-cutoff holography. They suggest that spectral methods offer a
natural language for describing holographic renormalization group flow
on non-compact boundaries and provide further evidence that the rich
mathematical structures underlying classical integrability continue to
play a fundamental role in the dynamics of gravitational systems away
from the asymptotic boundary.

The remainder of this paper is organized as follows.
Section~\ref{sec:three-dim-gravity} reviews the Chern--Simons
formulation of three-dimensional Einstein gravity together with the
diagonal reduction that gives rise to two independent chiral sectors.
In Section~\ref{sec:metric-formalism} we reconstruct the bulk metric
and discuss both its Fefferman--Graham expansion and its boundary ADM
decomposition. Section~\ref{sec:symplectic-structure} develops the
covariant symplectic structure and derives the current algebra, while
Section~\ref{sec:killing-symmetries} analyzes the associated Killing
symmetries together with the corresponding zero-mode charges.
Section~\ref{sec:fluid-gravity} establishes the exact fluid/gravity
dictionary and constructs the quasi-local Brown--York stress tensor at
finite radial position. In Section~\ref{sec:ttbar} we derive the exact
radial flow equation and demonstrate its interpretation as a {$\sqrt{T\bar{T}}$} deformation. Section~\ref{sec:vanishing-symplectic-flux}
formulates the Hamiltonian description of the boundary dynamics, and
Section~\ref{sec:bi-hamiltonian-structure} develops the associated
bi-Hamiltonian KdV-type hierarchy with arbitrary central charge. The Lax
representation, spectral problem, and inverse-scattering formulation,
including explicit soliton solutions, are presented in
Section~\ref{sec:lax-pair-formalism}. In
Section~\ref{sec:ttbar-deformed-kdv} we combine these ingredients to
investigate the finite-cutoff deformation of the integrable hierarchy,
establish the exact conservation and non-conservation laws for the
quasi-local observables. We conclude in
Section~\ref{sec:Discussion} with a discussion of the gravitational
interpretation of the spectral data, the significance of non-compact
boundaries, and several directions for future investigation. An outlook
is presented in Section~\ref{sec:Outlook}.

\section{Three-dimensional gravity}\label{sec:three-dim-gravity}

Einstein gravity in three spacetime dimensions with a negative cosmological constant admits an equivalent formulation as a Chern--Simons (CS) gauge theory~\cite{Achucarro:1987vz,Witten:1988hc}. This correspondence provides a powerful geometric and algebraic description of the theory, making manifest its topological character and greatly simplifying the analysis of its phase space, symmetries, and surface charges. In contrast to higher-dimensional gravity, pure three-dimensional Einstein gravity possesses no local propagating bulk degrees of freedom: the equations of motion constrain the spacetime to be locally maximally symmetric, and all nontrivial physical information is encoded in global data, boundary excitations, and holonomies around noncontractible cycles~\cite{Brown:1986nw,Carlip:1998uc}. 

For a negative cosmological constant $\Lambda = -1/\ell^2$, the Einstein--Hilbert action in first-order form can be written as the difference of two CS actions associated with independent $\mathfrak{sl}(2,\mathbb{R})$ gauge connections,
\begin{equation}\label{eq:EC-CS}
    S_{\text{\tiny EC}}[e,\omega]
    =
    S_{\text{\tiny CS}}[A^+]
    -
    S_{\text{\tiny CS}}[A^-] ,
\end{equation}
where
\begin{equation}\label{eq:Apm}
    A^\pm = \omega \pm \frac{1}{\ell} e .
\end{equation}
Here $e=e^m L_m$ denotes the dreibein, $\omega=\omega^m L_m$ is the dualized spin connection, and $L_m$ are generators of $\mathfrak{sl}(2,\mathbb{R})$. The equivalence between the metric and gauge-theoretic formulations follows from the isomorphism
\begin{equation}
    \mathfrak{so}(2,2)
    \cong
    \mathfrak{sl}(2,\mathbb{R})
    \oplus
    \mathfrak{sl}(2,\mathbb{R}),
\end{equation}
which realizes the AdS$_3$ isometry algebra as two commuting chiral sectors.

The CS action for a $\mathfrak{g}$-valued connection
$A \in \Lambda^1(\mathcal{M},\mathfrak{g})$
on a three-manifold $\mathcal{M}$ is
\begin{equation}\label{eq:CS-action}
    \begin{split}
        S_{\text{\tiny CS}}[A]
    = & \
    \frac{k}{4\pi}
    \int_{\mathcal{M}}
    \Big\langle
        A \wedge \dif A
        +
        \frac{2}{3} A \wedge A \wedge A
    \Big\rangle \\
 &   +
    \int_{\partial\mathcal{M}}
    \mathrm{L}_{\text{\tiny bdy}}[A] \, ,
    \end{split}
\end{equation}
where $k=\ell/(4G)$ is the CS level, $\langle \cdot,\cdot\rangle$ denotes a nondegenerate invariant bilinear form on $\mathfrak{g}$, and $\mathrm{L}_{\text{\tiny bdy}}$ is a boundary contribution chosen such that the variational principle is well defined for the boundary conditions under consideration~\cite{Banados:1994tn,Coussaert:1995zp}. The bulk action is gauge invariant up to a total derivative as a consequence of the Ad-invariance of the bilinear form.

Varying the action yields
\begin{equation}
    \delta S_{\text{\tiny CS}}
    =
    \frac{k}{2\pi}
    \int_{\mathcal{M}}
    \langle
        F \wedge \delta A
    \rangle
    +
    \int_{\partial\mathcal{M}}
    \Theta_{\text{\tiny CS}}[\delta A;A] ,
\end{equation}
where $ F = \dif A + A \wedge A$ is the curvature two-form and
\begin{equation}
    \Theta_{\text{\tiny CS}}[\delta A;A]
    =
    -\frac{k}{4\pi}
    \langle
        A \wedge \delta A
    \rangle
    +
    \delta \mathrm{L}_{\text{\tiny bdy}}[A]
\end{equation}
is the symplectic potential. The equations of motion are simply $ F=0$.

Consequently, all classical solutions are locally AdS$_3$, while globally inequivalent geometries arise from distinct holonomy sectors and boundary conditions~\cite{Banados:1992wn}.

A convenient basis for $\mathfrak{sl}(2,\mathbb{R})$ is provided by generators $L_m$ with
$m\in\{-1,0,1\}$ satisfying
\begin{equation}\label{eq:sl2alg}
    [L_m,L_n]
    =
    (m-n)L_{m+n}\, .
\end{equation}
The invariant bilinear form may be chosen as
\begin{equation}\label{eq:kappa}
    \kappa_{mn}
    =
    \langle L_m,L_n\rangle
    =
    \begin{pmatrix}
        0 & 0 & -1 \\
        0 & \frac12 & 0 \\
        -1 & 0 & 0
    \end{pmatrix}.
\end{equation}

To analyze the phase space it is convenient to employ a radial gauge decomposition~\cite{Banados:1998gg,Grumiller:2016pqb}
\begin{equation}\label{eq:radialgauge}
    A
    =
    b^{-1}(r)
    \left(
        \dif + a
    \right)
    b(r),
\end{equation}
where the group element $b(r)$ depends only on the radial coordinate and
\begin{equation}
    a = a_t \dif t + a_x \dif x \, ,
\end{equation}
is a connection defined on the boundary directions. In this gauge the radial dependence is pure gauge, and the nontrivial physical data are entirely encoded in the boundary fields contained in $a$.

With the gauge choice adopted throughout this work, the symplectic potential evaluated on boundary segments other than constant--$r$ hypersurfaces reduces to a total field-space variation and therefore does not contribute nontrivially to the presymplectic structure. In particular, these segments do not support independent dynamical degrees of freedom, as their associated contributions can be absorbed into boundary counterterms. Consequently, the only boundary component carrying nontrivial canonical information is the causal hypersurface at $r=\textit{const.}$, which encodes the effective boundary dynamics of the system. More precisely, the CS symplectic potential takes the form
\begin{equation}\label{S-P-C-C-01}
\begin{split}
\Theta_{\text{\tiny CS}}[\delta A;A]
=&
-\frac{k}{4\pi}
\left\langle
a \wedge \delta a
\right\rangle
\\
&+
\delta
\left(
\mathrm{L}_{\text{\tiny bdy}}[A]
-\frac{k}{4\pi}
\left\langle
\dif b\, b^{-1}\wedge a
\right\rangle
\right),
\end{split}
\end{equation}
where the second term is manifestly an exact variation in field space and hence does not affect the presymplectic form. The nontrivial contribution arises solely from the first term, which governs the induced phase space structure on the $r=\textit{const.}$ boundary. This feature is consistent with the standard treatment of boundary dynamics in CS formulations of gravity, where suitable gauge choices and boundary conditions isolate physical edge modes on distinguished boundary components while rendering other boundary contributions dynamically trivial.

{We may choose $\mathrm{L}{\text{\tiny bdy}}[A]$ such that the last term in \eqref{S-P-C-C-01} vanishes. In this formulation, $\mathrm{L}{\text{\tiny bdy}}[A]$ may be interpreted as a boundary counterterm whose role is twofold: it removes the divergent contribution to the symplectic potential and renders the resulting symplectic structure independent of the radial coordinate. Consequently, the symplectic structure reduces to a purely boundary contribution,
\begin{equation}
 \boldsymbol{\Theta}_{\text{\tiny CS}}
    =
    -\frac{k}{4\pi}
    \int_{\mathcal{B}}
    \langle
        a \wedge \delta a
    \rangle  
\end{equation}}
where $\mathcal{B}$ denotes the relevant causal boundary (located at $r=\textit{const.}$) component of spacetime.

In the following we focus on a diagonal subsector of the phase space~\cite{Afshar:2016uax,Grumiller:2019tyl,Ojeda:2019xih}, parametrized by
\begin{equation}\label{eq:diagsector}
    a_\pm
    =
    \pm \mathcal{A}^\pm L_0 ,
\end{equation}
with
\begin{equation}\label{eq:Apm-sector}
    \mathcal{A}^\pm
    =
    \mp 2 \mu_\pm(t,x)\,\dif t
    -
    \frac{2\pi}{k}
    \mathcal{J}_\pm(t,x)\,\dif x .
\end{equation}
The functions $\mu_\pm$ and $\mathcal{J}_\pm$ characterize the boundary data in the left- and right-moving sectors, respectively. Since the connection takes values entirely along the Cartan generator $L_0$, the nonlinear contribution $a\wedge a$ vanishes identically and the flatness condition reduces to an abelian equation. Substituting~\eqref{eq:Apm-sector} into $F=0$ gives
\begin{equation}\label{eq:J-eom}
    \partial_t \mathcal{J}_\pm
    =
    \pm
    \frac{k}{\pi}
    \partial_x \mu_\pm .
\end{equation}
This relation has the structure of a chiral continuity equation in two dimensions. In particular, $\mathcal{J}_\pm$ may be interpreted as chiral charge densities, while $\mu_\pm$ play the role of associated chemical potentials sourcing the corresponding currents. The dynamics in this reduced sector is therefore entirely governed by boundary transport equations rather than bulk propagation.

To ensure a well-defined phase space and finite surface charges, suitable boundary conditions must be imposed along the spatial direction. Two natural possibilities are:
\begin{enumerate}
    \item non-compact spatial slices with sufficiently rapid falloff at spatial infinity;
    \item periodic boundary conditions corresponding to cylindrical topology.
\end{enumerate}
In either case, one requires the integrated charges to remain finite,
\begin{equation}\label{eq:finitecharges}
    \left|
    \int_{\Sigma}
    \dif x\,
    \mathcal{J}_\pm(t,x)
    \right|
    < \infty ,
\end{equation}
where $\Sigma =
    \mathbb{R}$ or $
    \Sigma = [0,L)$. The choice of spatial topology and boundary conditions plays an essential role in determining the admissible symmetry algebra and the spectrum of physical states.

Residual gauge transformations preserving both the radial gauge~\eqref{eq:radialgauge} and the diagonal form~\eqref{eq:diagsector} are generated by
\begin{equation}
    \delta_\epsilon a
    =
    \dif \epsilon + [a,\epsilon],
    \qquad
    \epsilon_\pm
    =
    -2 \epsilon_\pm(t,x)L_0 \, .
\end{equation}
Their action on the boundary fields is
\begin{equation}\label{eq:gaugetransf}
    \delta_\epsilon \mathcal{J}_\pm
    =
    \pm
    \frac{k}{\pi}
    \partial_x \epsilon_\pm ,
    \qquad
    \delta_\epsilon \mu_\pm
    =
    \partial_t \epsilon_\pm .
\end{equation}
These transformations are improper gauge symmetries, namely gauge transformations with nonvanishing canonical generators acting nontrivially on the physical phase space~\cite{Regge:1974zd,Barnich:2001jy}. The arbitrary functions $\epsilon_\pm(t,x)$ generate an infinite-dimensional symmetry algebra whose centrally extended surface-charge realization depends on the chosen boundary conditions.

\paragraph{Generic solution.}

The field equations~\eqref{eq:J-eom} are first-order constraints that can be solved locally by introducing two scalar potentials $\Phi_\pm(t,x)$, one for each chiral sector. In this parametrization, the charge densities are expressed as total spatial derivatives,
\begin{equation}
\label{eq:J-to-Phi-clean}
    \mathcal{J}_\pm(t,x)
    =
    \frac{k}{\pi}\,\partial_x \Phi_\pm(t,x),
\end{equation}
which ensures integrability of the current algebra. Substituting~\eqref{eq:J-to-Phi-clean} into the equations of motion fixes the chemical potentials up to the expected chiral ambiguity. A consistent choice that solves~\eqref{eq:J-eom} identically is
\begin{equation}
\label{eq:mu-to-Phi-clean}
    \mu_\pm(t,x)
    =
    \pm \partial_t \Phi_\pm(t,x).
\end{equation}
With these definitions, the evolution equation becomes an identity, confirming that $\Phi_\pm$ provide a complete parametrization of the classical phase space in this sector.

The potentials $\Phi_\pm$ are not uniquely defined, since only their derivatives enter physical observables. This redundancy is precisely the remnant of the underlying gauge freedom. Indeed, under the residual transformations~\eqref{eq:gaugetransf}, one finds
\begin{equation}
\label{eq:Phi-transform}
    \delta_\epsilon \Phi_\pm
    =
    \pm \epsilon_\pm(t,x),
\end{equation}
so that $\Phi_\pm$ shift by arbitrary functions on the boundary. As a consequence, $\Phi_\pm$ behave as Stueckelberg-like fields.

This structure clarifies the nature of the reduced phase space: the physical degrees of freedom reside in equivalence classes of $\Phi_\pm$ modulo shifts, and all local observables are functionals of their derivatives. In particular, the symplectic form reduces to a boundary theory of chiral bosons, consistent with the appearance of affine current algebras in the canonical description of three-dimensional gravity with boundary degrees of freedom~\cite{Witten:1988hc,Banados:1994tn,Elitzur:1989nr}.

\section{Metric formalism}\label{sec:metric-formalism}

The bulk spacetime metric can be reconstructed from the CS formulation of three-dimensional gravity. Once a choice of radial group element $b^\pm(r)$ is made, the radial dependence of the gauge connections is fixed, and the spacetime geometry follows from the standard identification between CS variables and the dreibein.

The line element is obtained from the bilinear pairing of the frame fields,
\begin{equation}\label{line-element}
    \dif s^2 = 2 \langle e , e \rangle
    = 2 \kappa_{mn}\, e^m e^n ,
\end{equation}
where $e^m$ denotes the dreibein one-form and $\kappa_{mn}$ is the normalized Killing form of $\mathfrak{sl}(2,\mathbb{R})$.

We consider a spacetime equipped with a radial foliation by hypersurfaces of constant $r$. Let $n_\mu$ be the outward-pointing unit normal and define the induced metric
\begin{equation}
    q_{\mu\nu} = g_{\mu\nu} - n_\mu n_\nu .
\end{equation}

\paragraph{Fefferman--Graham gauge.}

A convenient radial gauge choice is implemented by the group element \cite{Ojeda:2019xih}
\begin{equation}
    b^\pm(r)
    =
    \exp\!\left[
    \pm \frac{1}{2}\ln\!\left(\frac{r}{\ell}\right)
    \left(L_{+1}-L_{-1}\right)
    \right],
\end{equation}
which realizes the Fefferman--Graham (FG) radial slicing for locally AdS$_3$ spacetimes.

In this gauge, the metric reconstructed from \eqref{line-element} assumes the form
\begin{equation}
    \dif s^2
    =
    \ell^2 \frac{\dif r^2}{r^2}
    + q_{ab}(r,x)\,\dif x^a \dif x^b ,
\end{equation}
with unit normal
\begin{equation}
    n_\mu \dif x^\mu = \ell \frac{\dif r}{r}.
\end{equation}

The induced metric admits the asymptotic structure
\begin{equation}\label{induced-metric-final}
    q_{ab}
    =
    \left(r^2 + \frac{\ell^4}{r^2}\right)\gamma_{ab}
    + \ell^2 \lambda_{ab},
\end{equation}
which is characteristic of locally AdS$_3$ geometries in FG coordinates. The tensor $\gamma_{ab}$ defines the conformal boundary data, while $\lambda_{ab}$ encodes subleading state-dependent contributions.

Explicitly, in terms of boundary fields $\mu_\pm$ and $\mathcal{J}_\pm$, one finds
\begin{equation}
\begin{aligned}
\gamma_{tt} &= -\mu_+\mu_-,
\qquad
\gamma_{tx} = \frac{\pi}{2k}(\mu_+\mathcal{J}_- - \mu_-\mathcal{J}_+),
\\
\gamma_{xx} &= \frac{\pi^2}{k^2}\mathcal{J}_+\mathcal{J}_-,
\qquad
\lambda_{tt} = \mu_+^2 + \mu_-^2,
\\
\lambda_{tx} &= \frac{\pi}{k}(\mu_+\mathcal{J}_+ - \mu_-\mathcal{J}_-),
\quad
\lambda_{xx} = \frac{\pi^2}{k^2}(\mathcal{J}_+^2 + \mathcal{J}_-^2).
\end{aligned}
\end{equation}

\paragraph{ADM decomposition on the boundary.}

We now consider a timelike boundary $\mathcal{B}$ with coordinates $x^a=(t,x)$ and induced metric $q_{ab}$. In two dimensions, any Lorentzian metric is locally conformally flat, and one may choose a representative $\sigma_{ab}$ in the same conformal class such that a $(1+1)$ ADM decomposition is available.

Writing $\sigma_{ab}$ in terms of a foliation by constant-$t$ slices yields
\begin{equation}\label{Boundary-metric-final}
    \dif \sigma^2
    =
    -N^2 \dif t^2
    + \left(\dif x + N^x \dif t\right)^2 ,
\end{equation}
where $N$ and $N^x$ denote the lapse and shift.

\paragraph{Orthonormal frame.}

An orthonormal frame adapted to \eqref{Boundary-metric-final} is given by
\begin{equation}
    u_a \dif x^a = -N \dif t \, ,
    \qquad
    s_a \dif x^a = \dif x + N^x \dif t\, ,
\end{equation}
with dual vectors
\begin{equation}
    u^a \partial_a = N^{-1}(\partial_t - N^x \partial_x) \, ,
    \qquad
    s^a \partial_a = \partial_x \, .
\end{equation}
These satisfy $u_a u^a=-1$, $s_a s^a=1$, and $u_a s^a=0$.

In two dimensions, the covariant derivatives of an orthonormal frame are fully determined by two scalars. One convenient decomposition is
\begin{subequations}
\begin{align}
    \nabla_a u_b &= \kappa_u\, u_a s_b - \kappa_s\, s_a s_b \, , \\
    \nabla_a s_b &= \kappa_u\, u_a u_b - \kappa_s\, s_a u_b \, ,
\end{align}
\end{subequations}
where $\kappa_u$ measures the acceleration of the timelike congruence and $\kappa_s$ encodes spatial variation of the slicing.

A direct computation yields
\begin{equation}
    \kappa_u = -\frac{\partial_x N}{N} \, ,
    \qquad
    \kappa_s = \frac{\partial_x N^x}{N}\, .
\end{equation}
Thus, the frame geometry is entirely controlled by spatial gradients of lapse and shift.

\paragraph{Geometric choice of lapse and shift.}

We now relate the boundary ADM data to the bulk solution by exploiting the conformal structure of \eqref{induced-metric-final}. Writing
\begin{equation}
    q_{ab} = W\, \sigma_{ab},
\end{equation}
the conformal factor is chosen as
\begin{equation}
    W =
    \frac{\pi^2}{k^2}
    \left(r \mathcal{J}_+ + \frac{\ell^2}{r}\mathcal{J}_-\right)
    \left(r \mathcal{J}_- + \frac{\ell^2}{r}\mathcal{J}_+\right).
\end{equation}

Matching with the ADM form uniquely determines lapse and shift:
\begin{subequations}\label{lapse-shift-final}
\begin{align}
    N &=
    \frac{\pi}{2k W}
    \left(r^2 - \frac{\ell^4}{r^2}\right)
    (\mu_+\mathcal{J}_- + \mu_-\mathcal{J}_+),
\\
    N^x &=
    \frac{\pi}{2k W}
    \Bigg[
        \left(r^2 + \frac{\ell^4}{r^2}\right)
        (\mu_+\mathcal{J}_- - \mu_-\mathcal{J}_+) \nonumber\\
        & \qquad\qquad\quad
        + 2\ell^2(\mu_+\mathcal{J}_+ - \mu_-\mathcal{J}_-)
    \Bigg].
\end{align}
\end{subequations}

For $r>\ell$, the foliation is well-defined and non-degenerate for generic boundary data. The surface $r=\ell$ corresponds only to a coordinate artifact of the chosen radial slicing.

\paragraph{Bulk geometry.}

Using the reconstruction map between CS fields and boundary variables, the metric can be written as
\begin{equation}\label{eq:ads3-metric-final}
    \begin{split}
    \dif s^2
    =&\;
    \ell^2 \frac{\dif r^2}{r^2}
    \\
    +&
    \ell^2
     \left(
        \frac{r}{\ell}\dif \Phi_+
        + \frac{\ell}{r}\dif \Phi_-
    \right)
    \left(
        \frac{r}{\ell}\dif \Phi_-
        + \frac{\ell}{r}\dif \Phi_+
    \right).
    \end{split}
\end{equation}

This is locally equivalent to the Ba\~nados family of solutions \cite{Banados:1998gg} and therefore satisfies the vacuum Einstein equations with negative cosmological constant. Different choices of $\Phi_\pm$ correspond to improper boundary diffeomorphisms generating inequivalent points in the phase space (boundary gravitons). The metric determinant is
\begin{equation}
    \sqrt{-g}
    =
    \frac{\ell}{2r^3}\left|r^4-\ell^4\right|.
\end{equation}
in coordinates $(r,\Phi_+,\Phi_-)$. The apparent degeneracy at $r=\ell$ is a coordinate singularity.

\section{Symplectic structure}\label{sec:symplectic-structure}

Restricting to the reduced diagonal sector, the symplectic potential and the associated symplectic two-form take a particularly simple form,
\begin{subequations}
\begin{align}
\boldsymbol{\Theta}
=&
- \int_{\mathcal{B}}
\left(
\mu_{+}\,\delta \mathcal{J}_{+}
+
\mu_{-}\,\delta \mathcal{J}_{-}
\right)
\,\dif t\,\dif x
\nonumber\\
&\quad
+
\frac{1}{2}\,
\delta
\int_{\mathcal{B}}
\left(
\mu_{+}\mathcal{J}_{+}
+
\mu_{-}\mathcal{J}_{-}
+
2\, \mathrm{L}^{r}_{\text{\tiny bdy}}
\right)
\,\dif t\,\dif x \, ,
\label{eq:symplectic-potential}
\\[0.2cm]
\boldsymbol{\Omega}
=&
-
\int_{\mathcal{B}}
\left(
\delta \mu_{+}\curlywedge \delta \mathcal{J}_{+}
+
\delta \mu_{-}\curlywedge \delta \mathcal{J}_{-}
\right)
\,\dif t\,\dif x \, .
\label{eq:symplectic-form}
\end{align}
\end{subequations}
The second term in~\eqref{eq:symplectic-potential} is an exact variation and therefore does not contribute to the symplectic structure. Consequently, the canonical pairs are given by $(\mu_{\pm},\mathcal{J}_{\pm})$, and the reduced phase space factorizes into two independent chiral sectors.

Within the covariant phase space formalism
\cite{Lee:1990nz,Iyer:1994ys,Wald:1999wa},
the variation of the surface charge associated with a gauge parameter $\epsilon$ is
\begin{equation}
\delta Q_{\text{\tiny CS}}(\epsilon)
=
\frac{k}{2\pi}
\int_{\Sigma}
\langle
\epsilon\,\delta a
\rangle ,
\label{eq:CS-charge-variation}
\end{equation}
where $\Sigma \subset \mathcal{B}$ denotes a constant-time section of the boundary. Since $\mathcal{B}$ is two-dimensional, $\Sigma$ is a codimension-two surface in the bulk spacetime. 

The charge variation evaluated at future and past temporal infinity differs by the bulk symplectic flux. More precisely,
\begin{equation}
\delta Q_{\text{\tiny CS}}(\epsilon)\Big|_{t\rightarrow +\infty}
=
\delta Q_{\text{\tiny CS}}(\epsilon)\Big|_{t\rightarrow -\infty}
+
\boldsymbol{\Omega}_{\text{\tiny CS}}
[\delta A,\delta_{\epsilon}A]\, .
\label{eq:charge-balance}
\end{equation}
Equation~\eqref{eq:charge-balance} therefore encodes the balance law for surface charges.

In the reduced phase space under consideration, the charge variation reduces to
\begin{equation}
\delta Q(\epsilon)
=
\int_{\Sigma}
\dif x
\left(
\epsilon_{+}\,\delta \mathcal{J}_{+}
+
\epsilon_{-}\,\delta \mathcal{J}_{-}
\right) .
\label{eq:reduced-charge}
\end{equation}
Assuming field-independent gauge parameters, the charges are integrable and one obtains
\begin{equation}
Q(\epsilon)
=
\int_{\Sigma}
\dif x
\left(
\epsilon_{+}\,\mathcal{J}_{+}
+
\epsilon_{-}\,\mathcal{J}_{-}
\right) .
\end{equation}

The algebra of surface charges can be computed from the equal-time Dirac brackets,
\begin{equation}
\begin{split}
\{Q(\epsilon_{1}),Q(\epsilon_{2})\}
 := & \ \delta_{\epsilon_{2}}Q(\epsilon_{1})\\
 = & \ Q([\epsilon_{1},\epsilon_{2}])
+
\mathcal{K}(\epsilon_{1},\epsilon_{2})
\end{split}
\label{eq:charge-algebra}
\end{equation}
where $\mathcal{K}(\epsilon_{1},\epsilon_{2})$ denotes the possible central extension. The corresponding equal-time current algebra takes the form
\begin{subequations}
\label{eq:current-algebra}
\begin{align}
\{\mathcal{J}_{\pm}(t,x),\mathcal{J}_{\pm}(t,y)\}
&=
\pm
\frac{k}{\pi}
\partial_{x}\delta(x-y) ,
\\
\{\mathcal{J}_{+}(t,x),\mathcal{J}_{-}(t,y)\}
&=
0 .
\end{align}
\end{subequations}
Equation~\eqref{eq:current-algebra} shows that the boundary dynamics is governed by two mutually commuting affine $\hat{\mathfrak{u}}(1)$ current algebras with opposite levels. The derivative of the Dirac delta function signals the presence of a non-trivial central extension, characteristic of Kac--Moody algebras arising from improper gauge transformations in CS theory.

The induced Poisson structure on the reduced phase space can be expressed in functional form as
\begin{equation}
\{F,G\}
=
\frac{k}{\pi}
\int_{\Sigma}
\dif x
\left(
\frac{\delta F}{\delta \mathcal{J}_{+}}
\partial_{x}
\frac{\delta G}{\delta \mathcal{J}_{+}}
-
\frac{\delta F}{\delta \mathcal{J}_{-}}
\partial_{x}
\frac{\delta G}{\delta \mathcal{J}_{-}}
\right) .
\label{eq:functional-poisson}
\end{equation}
This Poisson structure defines the reduced phase space of the theory and provides a canonical framework underlying the boundary dynamics. In particular, once the chemical potentials $\mu_{\pm}$ are specified as functionals of the currents, the resulting evolution equations define an integrable Hamiltonian flow generated by~\eqref{eq:functional-poisson}.

\section{Killing symmetries and zero modes}\label{sec:killing-symmetries}

The gauge transformation laws \eqref{eq:gaugetransf} admit a distinguished subclass of transformations generated by field-independent parameters. These correspond to rigid symmetries of the phase space and leave the dynamical configuration invariant. Specifically, by restricting the gauge parameters to constant values,
\begin{equation}\label{Killing-epsilon}
\epsilon_\pm
=
\mathrm{k}_\pm,
\qquad
\partial_t \mathrm{k}_\pm
=
\partial_x \mathrm{k}_\pm
=
0,
\qquad
\delta \mathrm{k}_\pm =0,
\end{equation}
the induced variations of the dynamical fields vanish. Hence these transformations generate isometries of the phase space and may be regarded as Killing symmetries.\footnote{The terminology is analogous to the notion of Killing vectors in gravitational theories, where the associated transformations preserve the background configuration. In the present context they correspond to rigid directions in the phase space along which the dynamical fields remain unchanged.}

The conserved charges associated with these rigid symmetries follow from the general surface-charge expression evaluated on constant parameters,
\begin{equation}
Q(\mathrm{k}_+,\mathrm{k}_-) =
\mathrm{k}_+H_0^+
+
\mathrm{k}_-H_0^-,
\label{Killing-charge}
\end{equation}
where
\begin{equation}\label{H0-boundary}
H_0^\pm
=
\int_\Sigma dx\,\mathcal{J}_\pm\, .
\end{equation}
denote the zero-mode charges of the chiral sectors. These quantities represent the global information carried by the field configuration.

The conservation of these charges follows directly from the equations of motion \eqref{eq:J-eom}. Taking a time derivative of \eqref{Killing-charge} yields
\begin{equation}
\partial_t
Q(\mathrm{k}_+,\mathrm{k}_-)
=
\frac{k}{\pi}
\left[
\mathrm{k}_+\mu_+
-
\mathrm{k}_-\mu_-
\right]^{x=+\infty}_{x=-\infty}\, .
\end{equation}
Consequently, the Killing charges are conserved whenever the boundary contribution vanishes. This condition is satisfied under admissible boundary conditions, such as sufficiently rapid decay of the fields and chemical potentials. Equivalently,
\begin{equation}
\partial_t H_0^\pm=0\, .
\end{equation}

Unlike generic gauge transformations generated by field-dependent and spatially varying parameters, Killing symmetries act uniformly throughout phase space and generate rigid global transformations. The corresponding charges probe only the zero-mode sector of the chiral fields and therefore capture global properties of the configuration rather than local dynamical excitations.

Moreover, the quantities $H_0^\pm$ characterize disconnected sectors of the phase space and act as labels of superselection sectors. Their physical interpretation depends on the global properties of the target space of $\Phi_\pm$. For non-compact target spaces, they assume continuous values and distinguish inequivalent asymptotic configurations.

\section{Fluid/Gravity Correspondence}\label{sec:fluid-gravity}

The Brown--York stress tensor on the timelike hypersurface
$\mathcal{B}$ admits a natural interpretation in terms of an
effective relativistic fluid propagating on the induced geometry
$(\mathcal{B},\sigma_{ab})$. Since the boundary is
two-dimensional, any symmetric rank-two tensor can be decomposed
with respect to an orthonormal basis $\{u^a,s^a\}$. Accordingly, the renormalized Brown--York tensor can be written as
\begin{equation}\label{BY-EMT-new}
T^{ab}
=
\rho\, u^a u^b
+
p\, s^a s^b
+
2J\, u^{(a}s^{b)} \, ,
\end{equation}
where $\rho$, $p$, and $J$ are scalar functions on $\mathcal{B}$.
These quantities are interpreted, respectively, as the local energy
density, pressure, and momentum density measured in the frame
defined by $u^a$. The trace of the stress tensor is therefore
\begin{equation}
T := \sigma_{ab}T^{ab} = -\rho + p \, .
\end{equation}

The decomposition \eqref{BY-EMT-new} is the most general form
compatible with the symmetries of a $(1+1)$-dimensional relativistic
system. In particular, the mixed term proportional to $J$
represents the energy flux along the spatial direction $s^a$,
or equivalently the momentum density carried by the effective fluid.

\paragraph{Conservation equations.}

The boundary dynamics is governed by the covariant conservation law
for the Brown--York tensor,
\begin{equation}\label{Conservation-equations-new}
\nabla_b T^{ab}=0 \, ,
\end{equation}
where $\nabla_a$ denotes the metric connection compatible
with $\sigma_{ab}$. Projecting along the vectors $u^a$ and $s^a$
yields two independent scalar equations. Using the ADM decomposition
of the boundary metric, one finds
\begin{subequations}\label{CEOM-new}
\begin{align}
\partial_t \rho
+
\partial_x (NJ-N^x\rho)
+ J\, \partial_x N
-  p \, \partial_x N^x
&=0 \, ,
\\
\partial_t J
+
\partial_x (Np-N^xJ)
+ \rho \,\partial_x N
- J \, \partial_x N^x
&=0 \, .
\end{align}
\end{subequations}
These equations take the form of relativistic hydrodynamic
conservation laws on a curved background geometry. The lapse
function $N$ and shift vector $N^x$ play the role of external
sources coupling to the local energy and momentum densities.

\paragraph{Symplectic structure and fluid variables.}

In terms of the decomposition \eqref{BY-EMT-new}, the boundary
contribution to the symplectic potential takes the form
\begin{equation}\label{SP-Hydro}
\begin{split}
\boldsymbol{\Theta}
&=
\int_{\mathcal{B}}
\left(
-\frac{1}{2}\sqrt{-\sigma}\,
T^{ab}\,
\delta \sigma_{ab}
+
\delta \mathrm{L}^{r}_{\text{\tiny bdy}}
\right)
\dif t\,\dif x
\\
&=
\int_{\mathcal{B}}
\left(
\rho\,\delta N
-
J\,\delta N^x
+
\delta \mathrm{L}^{r}_{\text{\tiny bdy}}
\right)
\dif t\,\dif x \, .
\end{split}
\end{equation}
The first contribution is the standard Brown--York symplectic
pairing between the induced metric and its conjugate tensor density,
while $\mathrm{L}^{r}_{\text{\tiny bdy}}$ denotes possible finite boundary
counterterms required for a well-defined variational principle and
finite quasi-local observables.

The fluid/gravity correspondence emerges by comparing the
gravitational symplectic potential \eqref{eq:symplectic-potential} with its
hydrodynamic form \eqref{SP-Hydro}. Imposing the bulk equations
of motion \eqref{eq:J-eom} together with the conservation law
\eqref{Conservation-equations-new}, one obtains an unambiguous
identification between the gravitational boundary data and the
fluid variables. In particular, the energy and momentum densities
are given by
\begin{subequations}
\begin{align}
\label{rho-final}
\rho
=&
\frac{\pi}{2k}
\left[
\frac{r^4+\ell^4}{r^4-\ell^4}
\left(
\mathcal{J}_+^2+\mathcal{J}_-^2
\right)
+
\frac{4\ell^2 r^2}{r^4-\ell^4}
\mathcal{J}_+\mathcal{J}_-
\right] ,
\\
\label{J-final}
J
=&
-\frac{\pi}{2k}
\left(
\mathcal{J}_+^2-\mathcal{J}_-^2
\right) .
\end{align}
\end{subequations}
A remarkable feature of these expressions is that the pressure
coincides with the energy density,
\begin{equation}\label{eos-stiff}
p=\rho \, .
\end{equation}
Consequently, the Brown--York tensor is traceless, $T = 0$,
which signals the emergence of an effective conformal fluid on the
boundary. This structure is consistent with the chiral decomposition
encoded by the currents $\mathcal{J}_\pm$ and with the conformal
nature of asymptotic dynamics in three-dimensional gravity.

{The finite subtraction scheme adopted here removes the state-independent finite-radius trace of the conventional Brown--York tensor, thereby yielding $T=0$ and $p=\rho$. We emphasize, however, that the resulting $\sqrt{T\bar T}$ characterization of the radial flow is tied to this particular renormalization scheme. In a different finite scheme, the stress tensor can acquire a nonzero state-independent trace, in which case the corresponding flow equation is modified accordingly.}

\paragraph{Hydrodynamic interpretation.}

The conservation equations \eqref{CEOM-new}, together with the
equation of state \eqref{eos-stiff}, define a relativistic
$(1+1)$-dimensional fluid system. Since $p=\rho$, the effective fluid is conformal and its speed of sound satisfies
\begin{equation}
c_s^2
=
\frac{\partial p}{\partial \rho}
=
1 \, .
\end{equation}
Thus, linear perturbations propagate along null directions of the
boundary geometry. This behavior is characteristic of
two-dimensional conformal hydrodynamics and reflects the absence
of an intrinsic dissipative scale in the present sector.

Moreover, the decomposition into chiral currents
$\mathcal{J}_\pm$ suggests an interpretation in terms of
left- and right-moving degrees of freedom propagating along the
boundary. The momentum density $J$ measures the imbalance between
the two sectors, while the energy density receives contributions
from both chiral components together with finite-radius mixing
terms proportional to $\mathcal{J}_+\mathcal{J}_-$.

\paragraph{Asymptotic regime.}

The fluid/gravity correspondence admits a smooth asymptotic limit
as $r\rightarrow\infty$. In this regime, the lapse and shift
functions behave as
\begin{subequations}
\begin{align}
N
&=
\frac{k}{2\pi}
\left(
\frac{\mu_+}{\mathcal{J}_+}
+
\frac{\mu_-}{\mathcal{J}_-}
\right)
+
\mathcal{O}(r^{-2}) \, ,
\\
N^x
&=
\frac{k}{2\pi}
\left(
\frac{\mu_+}{\mathcal{J}_+}
-
\frac{\mu_-}{\mathcal{J}_-}
\right)
+
\mathcal{O}(r^{-2}) \, .
\end{align}
\end{subequations}
Both quantities therefore approach finite, $r$-independent limits
at the conformal boundary. Expanding the energy density at large
$r$ gives
\begin{equation}
\rho
=
\frac{\pi}{2k}
\left(
\mathcal{J}_+^2+\mathcal{J}_-^2
\right)
+
\frac{2\pi\ell^2}{kr^2}
\mathcal{J}_+\mathcal{J}_-
+
\mathcal{O}(r^{-4}) \, ,
\end{equation}
while the momentum density remains intact. Consequently, the
Brown--York tensor admits a well-defined asymptotic limit, and
the associated fluid variables remain finite at the conformal
boundary. This ensures the consistency of the fluid/gravity map
in the asymptotic region and provides a quasi-local extension of
the standard holographic stress tensor construction.

\paragraph{Quasi-local energy and momentum.}

Given a constant-time spatial slice $\Sigma\subset\mathcal{B}$,
the quasi-local surface charges associated with the effective
fluid description are naturally defined as
\begin{subequations}\label{QLEM}
\begin{align}
\mathscr{E}(t,r)
&=
\int_{\Sigma}
\rho\,\dif x \, ,
\\
\mathscr{J}(t,r)
&=
\int_{\Sigma}
J\,\dif x \, .
\end{align}
\end{subequations}
These observables characterize the distribution of energy and
momentum on a finite radial hypersurface and therefore provide a
quasi-local probe of the bulk gravitational dynamics.

An important feature of the construction is the explicit radial
dependence of the quasi-local energy. The radial coordinate $r$
may be interpreted as parametrizing a family of effective boundary
theories defined on finite cutoff surfaces. From this viewpoint,
radial evolution naturally acquires the interpretation of a flow
in theory space, analogous to a renormalization group (RG) flow in the
dual field theory description (\emph{cf.} \cite{Brown:1992br,Balasubramanian:1999re,Bhattacharyya:2008xc,Hubeny:2011hd})

The interpretation of finite radial hypersurfaces as defining a one-parameter family of scale-dependent effective theories is naturally connected to the framework of the holographic RG \cite{deBoer:1999xf,Skenderis:2002wp,Bianchi:2001kw,Papadimitriou:2004ap}. 
In holographic setups, the radial coordinate plays the role of an energy scale in the dual field theory, such that radial evolution in the bulk geometrizes the RG flow of boundary observables. 
From this perspective, quasi-local quantities defined on finite radial slices should be regarded as running couplings or scale-dependent expectation values in the effective theory associated with the cutoff surface.

In the particular case of AdS$_3$/CFT$_2$ \cite{Brown:1986nw,Coussaert:1995zp,Achucarro:1987vz,Witten:1988hc,Strominger:1997eq,Maldacena:1997re, Witten:1998qj,Gubser:1998bc}, considerable evidence indicates that introducing a finite radial cutoff in the bulk corresponds to deforming the dual conformal field theory by the marginal composite operator $\sqrt{T\bar T}$ \cite{Rodriguez:2021tcz,Tempo:2022ndz,Babaei-Aghbolagh:2022uij,Babaei-Aghbolagh:2022leo,Conti:2022egv,Ferko:2022cix,Babaei-Aghbolagh:2024hti}. 
This deformation preserves integrability while modifying the ultraviolet behavior of the theory in a controlled manner. 
The resulting cutoff-dependent dynamics reproduces several characteristic features of finite-radius holography, including the deformation of the energy spectrum, modified thermodynamics, and the emergence of state-dependent effective geometries \cite{Guica:2019nzm,Hartman:2018tkw,Aharony:2018bad,Taylor:2018xcy}. 
In this correspondence, the radial position of the bulk hypersurface is identified with the energy scale at which the deformed theory is probed.

Accordingly, the radial evolution of quasi-local stress tensors, surface charges, and other boundary observables admits an interpretation as an RG flow driven by the finite cutoff. 
The associated flow equations encode how the effective theory changes under variations of the radial scale and provide a bridge between bulk gravitational dynamics and RG evolution in the dual description. 
This viewpoint is particularly useful in non-asymptotic settings, where finite-radius observables furnish a natural framework for characterizing the dynamics without taking the asymptotic boundary limit.

\section{\texorpdfstring{$\sqrt{T\bar T}$}{TTbar} deformation}
\label{sec:ttbar}

The interpretation of a finite radial cutoff in AdS$_3$ as an marginal deformation of the dual two-dimensional field theory by the operator $\sqrt{T\bar T}$ is by now well established in holography~\cite{Ebert:2023tih,Tempo:2022ndz,Rodriguez:2021tcz,Banerjee:2026qyc}. Most existing derivations are formulated either in terms of the finite-volume energy spectrum on a spatial circle or through the universal trace relation obeyed by the holographic stress tensor after the appropriate counterterms have been included. In these approaches, the radial evolution of the quasi-local stress tensor is usually extracted from a FG expansion and is therefore not presented as a closed local evolution equation written solely in terms of the stress tensor itself.

The formalism developed in this work provides precisely such a description. The quasi-local fluid variables $\rho$ and $J$, introduced in Sec.~\ref{sec:fluid-gravity}, are known exactly as functions of the radial coordinate $r$ and the $r$-independent chiral data $\mathcal{J}_\pm$, see Eqs.~\eqref{rho-final}--\eqref{J-final}. Unlike the conventional near-boundary expansion, these expressions are valid to all orders in the radial coordinate and for arbitrary boundary profiles $\mathcal{J}_\pm(t,x)$. As a consequence, the chiral variables can be eliminated algebraically, yielding a closed differential equation governing the radial evolution of the quasi-local stress tensor without any explicit reference to the underlying bulk geometry.

It is important to emphasize the scope of this construction. Our analysis establishes an exact local flow equation relating the radial variation of the quasi-local energy density to an invariant quadratic combination of the stress tensor. {By itself, however, this result does not imply a corresponding flow equation for finite-volume energy levels. In particular, it should not be interpreted as establishing a universal spectral flow analogous to that arising in the ordinary $T\bar T$ deformation~\cite{McGough:2016lol,Smirnov:2016lqw}. Establishing such a relation would require, in addition, a well-defined identification of the finite-volume spectrum and its dependence on the deformation parameter, which is beyond the scope of the present analysis.} Such an identification requires a precise dictionary between the radial coordinate and the deformation parameter, together with a matching between local quasi-local observables and global spectral quantities. The discussion below should therefore be viewed as an exact statement about the radial evolution in the present holographic framework, together with its relation to the standard $T\bar T$ picture.

Differentiating the energy density $\rho$ with respect to the radial coordinate while keeping the boundary data fixed, and subsequently eliminating $\mathcal{J}_\pm$ in favor of $\rho$ and $J$, one finds
\begin{equation}\label{eq:ttbar-flow-r}
    \partial_r \rho
    =
    -\frac{4\ell^2 r}{r^4-\ell^4}\,
    \sqrt{\rho^2-J^2}\, .
\end{equation}
The momentum density \eqref{J-final} is independent of the radial cutoff, $\partial_r J=0$.

A straightforward computation further shows that the quantity appearing under the square root in \eqref{eq:ttbar-flow-r} is identically a perfect square for arbitrary $\mathcal{J}_\pm$ throughout the physical region $r>\ell$. Consequently, choosing the principal branch, the square root is everywhere real along the radial flow, and the evolution equation remains manifestly real over the entire domain of interest.

The quadratic combination $\rho^2-J^2$ naturally admits an invariant interpretation. Expressing the Brown--York stress tensor in the orthonormal frame $(u,s)$,
\begin{equation}
    T_{\text{\tiny (frame)}}=
    \begin{pmatrix}
        T_{uu} & T_{us}\\
        T_{us} & T_{ss}
    \end{pmatrix},
\end{equation}
its determinant is
\begin{equation}
    \det T_{\text{\tiny (frame)}}=\rho^2-J^2\, .
\end{equation}
Since the above expression is evaluated in an orthonormal frame, it represents an invariant quadratic scalar constructed from the quasi-local stress tensor rather than a consequence of a particular coordinate choice or normalization.

It is convenient to introduce the dimensionless radial parameter $\varepsilon=\ell^2/r^2$, which vanishes at the asymptotic AdS boundary and increases monotonically toward the coordinate locus $r=\ell$. Rewriting \eqref{eq:ttbar-flow-r} in terms of $\varepsilon$ gives
\begin{equation}
    \partial_\varepsilon T_{uu}
    =
    \frac{2 }{1-\varepsilon^4}\,
    \sqrt{\det T_{\text{\tiny (frame)}} }\, .
\end{equation}
This form makes the flow particularly transparent: the radial evolution is governed entirely by the determinant of the quasi-local stress tensor, closely paralleling the structure characteristic of holographic {$\sqrt{T\bar{T}}$} deformations.

\section{Vanishing symplectic flux}\label{sec:vanishing-symplectic-flux}

The boundary dynamics induced by the flatness condition depends crucially on the boundary conditions imposed at the asymptotic causal boundary. Within the covariant phase space framework, a consistent variational principle requires that no symplectic current flows through the boundary. Equivalently, the symplectic flux through the asymptotic boundary must vanish,
\begin{equation}\label{V-S-F}
    \boldsymbol{\Omega}=0 \, .
\end{equation}
This condition guarantees that the presymplectic structure remains conserved under deformations of the hypersurface and therefore defines a genuine symplectic structure on the reduced phase space. Physically, it ensures that the system behaves as a closed Hamiltonian system, with no exchange of symplectic charge with the exterior region.

In the present context, the vanishing-flux condition implies the integrability of the chemical potentials $\mu_\pm$ in field space. Consequently, at least locally on the phase space, these quantities may be represented as functional derivatives of a Hamiltonian functional constructed from the canonical boundary fields. This relation plays a central role in establishing a Hamiltonian description of the boundary dynamics.


The reduced phase space is parametrized by the two dynamical fields
$\mathcal{J}_\pm$, whose evolution is governed by the equations of motion \eqref{eq:J-eom}. One may therefore introduce a local Hamiltonian functional of the form
\begin{equation}\label{H-Functional}
H[\mathcal{J}_+,\mathcal{J}_-]
=
\int_{\Sigma}\mathrm{d}x\,
\mathcal{H}
\big(
\mathcal{J}_\pm,
\partial_x\mathcal{J}_\pm,
\ldots
\big),
\end{equation}
where locality means that the Hamiltonian density $\mathcal{H}$ depends on the fields and only finitely many of their spatial derivatives. Different choices of $\mathcal{H}$ correspond to different admissible boundary conditions and consequently define different boundary theories.

The requirement of functional differentiability of $H$, together with the vanishing symplectic flux condition \eqref{V-S-F}, implies that the chemical potentials are determined by the Euler--Lagrange derivatives of the Hamiltonian functional,
\begin{equation}\label{mu-E-L}
\mu_\pm
=
\frac{\delta H}{\delta \mathcal{J}_\pm}.
\end{equation}
Hence the chemical potentials act as conjugate sources associated with the canonical variables and encode the boundary evolution generated by the Hamiltonian flow.

Substituting \eqref{mu-E-L} into the symplectic potential \eqref{eq:symplectic-potential}, one finds that, modulo total derivative terms and exact variations in field space, the symplectic potential becomes itself an exact variation. Accordingly, one may choose a representative in the corresponding equivalence class such that the reduced dynamics is encoded by the effective boundary Lagrangian density
\footnote{
Using the definitions of the canonical energy and momentum densities,
\begin{equation*}
\frac{1}{2}
\left(
\mu_+\mathcal{J}_+
+
\mu_-\mathcal{J}_-
\right)
=
N\rho
-
N^xJ \, .
\end{equation*}
}
\begin{equation}\label{Lbdy}
\mathrm{L}^{r}_{\text{\tiny bdy}}
=
\mathcal{H}
-
N\rho
+
N^xJ .
\end{equation}
The resulting boundary theory therefore admits a Hamiltonian formulation in which the symplectic structure is inherited directly from the reduced phase space construction. In this description, the choice of boundary conditions is entirely encoded in the Hamiltonian density $\mathcal{H}$.

A natural choice for the improper gauge parameters is to identify them with the chemical potentials,
\begin{equation}
\epsilon_\pm=\mu_\pm .
\end{equation}
With this identification, the variation of the surface charge \eqref{eq:reduced-charge} becomes integrable and reproduces the Hamiltonian generating the dynamics,
\begin{equation}
H
=
Q(\mu_+,\mu_-).
\end{equation}
This relation reflects a standard feature of the covariant phase space approach: time evolution is generated by a particular improper gauge transformation whose parameter is fixed by the boundary sources. The Hamiltonian thus emerges as the conserved charge associated with the symmetry generator implementing time translations compatible with the chosen boundary conditions.

The existence of a well-defined Hamiltonian generator follows from the absence of symplectic leakage through the boundary. Moreover, the Hamiltonian functional carries no explicit time dependence, its value is preserved along the Hamiltonian flow,
\begin{equation}
\frac{\mathrm{d}H}{\mathrm{d}t}=0\, .
\end{equation}
This conservation law expresses the invariance of the corresponding surface charge under the symmetry flow generated by the Hamiltonian itself.

Once the Hamiltonian functional
$H[\mathcal{J}_+,\mathcal{J}_-]$
is specified, the equations of motion acquire a canonical form in terms of the Poisson structure on the reduced phase space. The evolution equations become
\begin{equation}\label{evolution-eq-J}
\partial_t\mathcal{J}_\pm
=
\left\{
\mathcal{J}_\pm,
H
\right\},
\end{equation}
where the Poisson bracket is defined as in
\eqref{eq:functional-poisson}. This equation describes the Hamiltonian flow generated by $H$ and is equivalent to the original equations of motion.

\section{Bi-Hamiltonian structure}\label{sec:bi-hamiltonian-structure}

We now assume that the Hamiltonian introduced in the previous section is well-defined and denote it by $H_{\rm P}:=H$. Suppose further that there exists another Hamiltonian functional
$H_{\rm D}$ such that the equations of motion admit two inequivalent Hamiltonian descriptions,
\begin{equation}\label{biHam}
    \pm\frac{\pi}{k}\,\partial_t \mathcal J_{\pm}
    =
    \mathrm D_{\pm}
    \frac{\delta H_{\rm D}}
         {\delta \mathcal J_{\pm}}
    =
    \mathrm P_{\pm}
    \frac{\delta H_{\rm P}}
         {\delta \mathcal J_{\pm}}\, .
\end{equation}
The existence of two compatible Hamiltonian formulations is the defining characteristic of a bi-Hamiltonian system and constitutes the basic mechanism underlying the generation of integrable hierarchies
\cite{magri1978simple,faddeev1987hamiltonian,olver1993applications}.

The operators
$\mathrm P_{\pm}$ and $\mathrm D_{\pm}$ are skew-adjoint differential operators acting on the phase space of fields
$\mathcal J_\pm$, defining Poisson brackets through \eqref{eq:functional-poisson}.

For the hierarchy considered here, the first Hamiltonian structure is chosen as $\mathrm P_\pm=\partial_x$, corresponding to the standard translationally invariant Poisson bracket. The second Hamiltonian structure is taken to be
\begin{equation}\label{Dpm-new}
    \mathrm D_\pm
    =
    \partial_x\mathcal J_\pm
    +
    2\mathcal J_\pm\partial_x
    -
    \frac{c_\pm}{24\pi}
    \partial_x^3\, .
\end{equation}
This operator is recognized as the second Gel'fand--Dikii Hamiltonian structure, or equivalently the classical Virasoro Poisson operator
\cite{Gelfand:1975rn,Dickey:1991xa}.

A fundamental requirement is the compatibility of the two Poisson structures. This means that the Schouten bracket of the corresponding Poisson tensors vanishes, which is equivalent to requiring that every linear combination
\begin{equation}
    \mathrm D_\pm^{(\alpha)}
    :=
    \mathrm D_\pm
    +
    \alpha\,
    \mathrm P_\pm,
    \qquad
    \alpha\in\mathbb R,
\end{equation}
again defines a valid Poisson operator. Compatibility guarantees the existence of infinitely many commuting Hamiltonian flows and therefore an associated integrable hierarchy.

\paragraph{Lenard recursion.}

The hierarchy of conserved quantities can be generated recursively by means of the Lenard--Magri scheme
\cite{magri1978simple,lax1968integrals}. Introducing a sequence of differential polynomials
$\mathcal R_I[\mathcal J_\pm]$, the recursion relation takes the form
\begin{equation}\label{Lenard-new}
    \mathrm D_\pm
    \mathcal R_I
    [\mathcal J_\pm]=\mathrm P_\pm
    \mathcal R_{I+1}
    [\mathcal J_\pm]\, .
\end{equation}
Starting from the seed $\mathcal R_0=1$,
the recursion generates the entire sequence of generalized Gel'fand--Dikii polynomials. The resulting functions are defined modulo elements in the kernel of
$\mathrm P_\pm$, which do not contribute to the associated conserved charges under standard boundary conditions. \footnote{The first members of the hierarchy are
\begin{equation*}
\begin{aligned}
\mathcal R_0
&=
1\, ,
\\
\mathcal R_1
&=
\mathcal J_\pm\, ,
\\
\mathcal R_2
&=
\frac32
\mathcal J_\pm^2
-
\frac{c_\pm}{24\pi}
\mathcal J_\pm''\, ,
\\
\mathcal R_3
&=
\frac52
\mathcal J_\pm^3
-
\frac{5c_\pm}{24\pi}
\mathcal J_\pm
\mathcal J_\pm''
-
\frac{5c_\pm}{48\pi}
(\mathcal J_\pm')^2
+
\frac{c_\pm^2}
     {576\pi^2}
\mathcal J_\pm''''\, .
\end{aligned}
\end{equation*}
}
\paragraph{Conserved charges.}

We next examine the integrability of the surface-charge variations \eqref{eq:reduced-charge}. 
Integrability requires that the one-form in field space be closed, that is $\delta^2Q(\epsilon)=0$, or equivalently,
\begin{equation}\label{integrability-new}
    \int_\Sigma
    \dif x
    \,
    \left(
    \delta\epsilon_+
    \curlywedge
    \delta\mathcal J_+
    +
    \delta\epsilon_-
    \curlywedge
    \delta\mathcal J_-
    \right)
    =
    0\, .
\end{equation}
A natural choice adapted to the hierarchy is
\begin{equation}\label{eps-choice-new}
    \epsilon_\pm
    =
    \mathcal R_I
    [\mathcal J_\pm]\, .
\end{equation}
This choice ensures that the corresponding one-form in field space is closed. Hence the charge variations
\begin{equation}
    \delta H_I^\pm
    =
    \int_\Sigma
    \dif x\,
    \mathcal R_I
    [\mathcal J_\pm]
    \delta\mathcal J_\pm
\end{equation}
are integrable. The corresponding Hamiltonians may be reconstructed using the homotopy operator in field space,~\footnote{The first few Hamiltonians are
\begin{equation*}
\begin{aligned}
H_0^\pm
&=
\int_\Sigma
\dif x\,
\mathcal J_\pm\, ,
\\
H_1^\pm
&=
\int_\Sigma
\dif x\,
\left[\frac{1}{2}\mathcal J_\pm^2 \right]\, ,
\\
H_2^\pm
&=
\int_\Sigma
\dif x
\left[
\frac12
\mathcal J_\pm^3
+
\frac{c_\pm}{48\pi}
(\mathcal J_\pm')^2
\right] ,\\
H_3^\pm
    &=
    \int_\Sigma \dif x \left[
        \frac{5}{8}\,\mathcal{J}_\pm^4
        +
        \frac{5c_\pm}{48\pi}\,\mathcal{J}_\pm (\mathcal{J}_\pm')^2
        +
        \frac{c_\pm^2}{1152\pi^2}\,(\mathcal{J}_\pm'')^2
    \right]\, .
\end{aligned}
\end{equation*}
}
\begin{equation}\label{Hamiltonian_homotopy}
    H_I^\pm
    =
    \int_0^1
    \dif s
    \int_\Sigma
    \dif x
    \,
    \mathcal R_I
    [s\mathcal J_\pm]
    \mathcal J_\pm\, .
\end{equation}
The Lenard relation immediately implies the Magri identity,
\begin{equation}
    \{H_{I+1}^\pm,H_J^\pm\}
    =
    \{H_I^\pm,H_{J+1}^\pm\}\, .
\end{equation}
Repeated application of this relation together with the antisymmetry of the Poisson bracket yields
\begin{equation}
    \{H_I^\pm,H_J^\pm\}
    =
    0,
    \qquad
    \forall I,J\ge0\, .
\end{equation}
Hence the Hamiltonians form an infinite family of mutually commuting conserved quantities. Each Hamiltonian flow preserves all other charges, establishing the \emph{Liouville integrability} of the hierarchy.

\paragraph{Anisotropic scaling symmetry.}

The integrable hierarchy generated by the conserved Hamiltonians \eqref{Hamiltonian_homotopy} possesses a homogeneous anisotropic scaling symmetry of Lifshitz type, a characteristic feature of many integrable systems and generalized KdV hierarchies \cite{lifshitz1941theory,dickey2003soliton,olver1993applications}. Under this transformation, spatial and temporal coordinates scale with different weights according to
\begin{equation}
\label{eq:Lifshitz_scaling_refined}
    x \;\rightarrow\; \zeta^{-1}x,
    \qquad
    t
    \;\rightarrow\;
    \zeta^{-(2I+1)}
    t \, ,
\end{equation}
where the integer $I$ labels the flow in the hierarchy. The dynamical exponent associated with the $I$-th flow is therefore $2I+1$, which reproduces the characteristic odd-order dispersion relation of generalized KdV-type systems.

The fields transform homogeneously as $\mathcal{J}_{\pm}
    \rightarrow\
    \zeta^{2}\mathcal{J}_{\pm}$,
while the Gel'fand--Dikii polynomials
$\mathcal{R}_{I}[\mathcal{J}_{\pm}]$ scale according to
\begin{equation}
\label{R_scaling}
    \mathcal{R}_{I}[\mathcal{J}_{\pm}]
    \;\rightarrow\;
    \zeta^{2I}
    \mathcal{R}_{I}[\mathcal{J}_{\pm}] \, .
\end{equation}
These scaling assignments follow from the requirement that the associated Lax operator transform homogeneously. 

Consequently, the Hamiltonian densities possess scaling dimension $2I+2$, and therefore the conserved Hamiltonians transform as
\begin{equation}
\label{Hamiltonian_scaling}
    H_{I}^\pm
    \;\rightarrow\;
    \zeta^{2I+1}
    H_{I}^\pm \, .
\end{equation}
The factor of $2I+1$ arises from the interplay between the scaling of the Hamiltonian density and that of the spatial integration measure,
$\dif x\rightarrow \zeta^{-1}\dif x$.
Hence each Hamiltonian carries a well-defined homogeneous degree under the Lifshitz scaling symmetry.

This anisotropic scaling structure is not merely a dimensional property but reflects the organization of the hierarchy itself: successive Hamiltonian flows are naturally ordered according to increasing scaling weight. In particular, higher members of the hierarchy correspond to progressively larger dynamical exponents and describe higher-derivative evolution equations with increasingly dispersive behavior.

\section{Lax pair formalism}\label{sec:lax-pair-formalism}

A central characterization of integrable systems is provided by the Lax formulation, in which nonlinear evolution equations are represented as isospectral deformations of auxiliary linear operators \cite{lax1968integrals,ablowitz1981solitons,dickey2003soliton}.  
This framework not only provides a compact description of the dynamics but also naturally explains the existence of infinitely many conserved quantities and the emergence of exact solution-generating techniques such as the inverse scattering transform.

Let $\mathcal{J}_{\pm}(t,x)$ denote a pair of dynamical fields evolving according to \eqref{biHam}. We assume that the dynamics can be represented through a pair of operators $(\mathrm{D}_{\pm},\mathrm{S}_{\pm})$ satisfying the Lax equation
\begin{equation}\label{Lax-eq}
    \partial_t \mathrm{S}_{\pm}
    =
    \pm [\mathrm{D}_{\pm},\mathrm{S}_{\pm}]\, .
\end{equation}
The commutator structure implies that the evolution of $\mathrm{S}_{\pm}$ is generated by a similarity transformation and therefore preserves its spectrum.

To make this statement explicit, consider the associated spectral problem
\begin{equation}
\label{Sch-eq}
    \mathrm{S}_{\pm}\psi_{\pm}
    =
    \lambda_{\pm}^{2}\psi_{\pm},
\end{equation}
where $\lambda_{\pm}\in\mathbb{C}$ is the spectral parameter and $\psi_{\pm}$ denotes the corresponding eigenfunction. Differentiating \eqref{Sch-eq} with respect to time and using \eqref{Lax-eq} gives
\begin{equation}
    \mathrm{S}_{\pm}
    \bigl(
        \partial_t\psi_{\pm}
        \mp
        \mathrm{D}_{\pm}\psi_{\pm}
    \bigr)
    =
    \lambda_{\pm}^{2}
    \bigl(
        \partial_t\psi_{\pm}
        \mp
        \mathrm{D}_{\pm}\psi_{\pm}
    \bigr)
   \, ,
\end{equation}
provided that $\partial_t\lambda_{\pm}=0$. Hence the evolution generated by \eqref{Lax-eq} preserves the spectrum of $\mathrm{S}_{\pm}$, and the flow is therefore isospectral.

The isospectral property is the underlying mechanism responsible for the existence of infinitely many conserved quantities. Indeed, spectral invariants of $\mathrm{S}_{\pm}$ remain constant under the evolution. Typical examples include traces of powers of $\mathrm{S}_{\pm}$ or, more generally, coefficients appearing in the asymptotic expansion of its resolvent. Such quantities generate the infinite tower of conserved charges characterizing integrable hierarchies \cite{faddeev1987hamiltonian,dickey2003soliton}.

\paragraph{Schr\"odinger operator.}

For the class of systems considered here, the Lax operator takes the form of a one-dimensional Schr\"odinger operator,
\begin{equation}
\label{Sch-operator}
    \mathrm{S}_{\pm}
    =
    -\partial_x^2
    +
    \frac{2}{\nu_\pm}\,
    \mathcal{J}_{\pm},
\end{equation}
where $\mathcal{J}_{\pm}(t,x)$ plays the role of a time-dependent potential and $\nu_\pm$ are some constants.

The associated integrable hierarchy is encoded in the isospectral deformations of \eqref{Sch-operator}. A standard procedure for extracting the hierarchy consists of introducing the logarithmic derivative
\begin{equation}
    \Gamma_{\pm}(t,x;\lambda)
    =\partial_x \ln{|\psi_{\pm}|},
\end{equation}
which transforms the spectral problem \eqref{Sch-eq} into the Riccati equation
\begin{equation}
\label{Riccati}
    \partial_x \Gamma_{\pm}
    +
    \Gamma_{\pm}^{2}
    -
    \frac{2}{\nu_\pm}\,
    \mathcal{J}_{\pm}
    +
    \lambda_{\pm}^{2}=0 \, .
\end{equation}

\paragraph{Spectral structure.}

The Lax representation also provides a natural description of the solution space through spectral methods. The precise structure of this space depends crucially on the choice of spatial geometry and boundary conditions.

At fixed time, \eqref{Sch-eq} defines a linear second-order ordinary differential equation in the spatial coordinate $x$. Consequently, its solution space is two-dimensional. Let
\begin{equation}
    \psi_{\pm}^{\text{ \tiny (1)}}
    (t,x;\lambda_{\pm}),
    \qquad
    \psi_{\pm}^{\text{ \tiny (2)}}
    (t,x;\lambda_{\pm}),
\end{equation}
be a pair of linearly independent solutions forming a fundamental basis. Any solution can be written as a linear combination of these basis elements. The associated Wronskian,
\begin{equation}
    W_{\pm}(t,\lambda_{\pm})
    =
    \psi_{\pm}^{\text{ \tiny (1)}}
    \partial_x
    \psi_{\pm}^{\text{ \tiny (2)}}
    -
    \psi_{\pm}^{\text{ \tiny (2)}}
    \partial_x
    \psi_{\pm}^{\text{ \tiny (1)}},
\end{equation}
is independent of $x$, as follows directly from the second-order nature of the differential equation.

For non-compact spatial slices, the problem becomes a scattering problem for the Schr\"odinger operator with potential $\mathcal{J}_{\pm}$ \cite{ablowitz1981solitons,faddeev1987hamiltonian}. The spectrum generally contains a continuous component with $\lambda_{\pm}^{2}\geq0$ and, depending on the form of the potential, a discrete set of bound states at negative energies. The spectral data are characterized by the Jost solutions and their associated scattering coefficients, including reflection and transmission amplitudes together with norming constants.

The full nonlinear dynamics can be reconstructed through the inverse scattering transform. In this picture, the discrete part of the spectrum is associated with solitonic excitations, whereas the continuous sector describes dispersive modes propagating on the background geometry.

\subsection{Solution space}

We consider the one-dimensional Schr\"odinger spectral problem
\eqref{Sch-eq}, defined by the differential operator
\eqref{Sch-operator}. For a fixed potential $\mathcal{J}_\pm(t,x)$, the
\emph{direct scattering problem} consists in determining the spectral data
associated with the eigenvalue equation, while the \emph{inverse problem}
asks whether $\mathcal{J}_\pm$ can be uniquely reconstructed from this data.
Under standard short-range assumptions, this reconstruction is achieved
non-perturbatively via the Gelfand--Levitan--Marchenko (GLM) formalism
\cite{marchenko2013sturm, faddeev1987hamiltonian, ablowitz1981solitons}.

\paragraph{Free reference problem.}

To set up the scattering framework, it is convenient to introduce the
reference problem obtained by switching off the potential. For
$\mathcal{J}_\pm=0$, the spectral equation reduces to
\begin{equation}
\label{eq:free_sch_clean}
    \partial_x^2 \phi_\pm + \lambda_\pm^2 \phi_\pm = 0,
\end{equation}
with $\lambda_\pm \in \mathbb{C}$. A convenient fundamental system is
\begin{equation}\label{S-S-Free}
\phi_\pm(t,x;\lambda_\pm)
=
\mathcal{C}_\pm(t;\lambda_\pm)\, e^{i\lambda_\pm x}\, ,
\qquad
\overline{\phi_\pm(t,x;\lambda_\pm)} \, ,
\end{equation}
where $\mathcal{C}_\pm$ encodes normalization freedom and possible time dependence dictated by the Lax evolution
\eqref{Lax-eq}. These free solutions define the asymptotic basis
for the interacting problem.

\paragraph{Jost solutions.}

We assume that $\mathcal{J}_\pm(t,x)$ are short-range in $x$, ensuring
that standard one-dimensional scattering theory applies; in particular,
$\mathcal{J}_\pm(t,x)$ is assumed integrable (or sufficiently rapidly decaying)
as $x \to \pm\infty$.

The Jost solutions $\psi_\pm(t,x;\lambda_\pm)$ are defined by the
asymptotic matching conditions
\begin{equation}
    \psi_\pm(t,x;\lambda_\pm)
    \sim
    \phi_\pm(t,x;\lambda_\pm),
    \qquad x \to \pm\infty \, .
\end{equation}
Under the above assumptions, the Jost solutions exist uniquely and are
analytic in $\lambda_\pm$ in the upper/lower half spectral planes,
depending on the chosen boundary conditions.

A key structural fact is that $\psi_\pm$ admit Volterra-type integral
representations
\begin{equation}
\label{eq:jost_volterra_clean}
\begin{split}
    \psi_\pm(t,x;\lambda_\pm)
=&\
\phi_\pm(t,x;\lambda_\pm)
\\
&+
\int_{\pm\infty}^{x}
K_\pm(t,x,y)\,
\phi_\pm(t,y;\lambda_\pm)\,\dif y,
\end{split}
\end{equation}
where the transformation kernel $K_\pm$ is independent of $\lambda_\pm$ and
satisfies the support conditions
\begin{subequations}
\begin{align}
K_+(t,x,y) &= 0 \qquad \text{for } y > x,\\
K_-(t,x,y) &= 0 \qquad \text{for } y < x.
\end{align}
\end{subequations}

The Volterra structure ensures existence and uniqueness of the Jost
solutions without any perturbative assumption. It also guarantees that the
analytic properties of $\psi_\pm$ are inherited from the free solutions
through a triangular integral operator.

\paragraph{Equation for the transformation kernel.}

Substituting \eqref{eq:jost_volterra_clean} into the Schr\"odinger
equation \eqref{Sch-eq} and using the free equation
\eqref{eq:free_sch_clean}, one obtains a second-order partial differential
equation for the transformation kernel:
\begin{equation}
\label{eq:kernel_pde_clean}
\left(
\partial_x^2 - \partial_y^2
\right) K_\pm(t,x,y)
=
\frac{2}{\nu_\pm}\,
\mathcal{J}_\pm(t,x)\, K_\pm(t,x,y).
\end{equation}

This relation holds in the domain compatible with the Volterra support
conditions. In particular, evaluating at coincident points $y=x$ yields a
local reconstruction formula
\begin{equation}\label{reconstruction-formula}
    \mathcal{J}_\pm(t,x)
    =
   \nu_\pm \,\partial_x K_\pm(t,x,x),
\end{equation}
which shows that the potential is encoded in the diagonal behavior of the
transformation kernel. In the inverse problem, $K_\pm$ is itself determined
by the scattering data through the GLM equation, so this relation provides
the final step in the reconstruction of $\mathcal{J}_\pm$.

It is important to note that the diagonal value of the kernel,
$K_\pm(t,x,y)$ evaluated at coincident points, determines the field $\Phi_\pm$ only up to an additive function of time. More precisely,
one may write
\begin{equation}
    \Phi_\pm(t,x)
    =
   \frac{\pi \nu_\pm}{k} K_\pm(t,x,x)
    +
    \varphi_\pm(t),
\end{equation}
where $\varphi_\pm(t)$ is an arbitrary function depending solely on
time. The appearance of this undetermined contribution reflects the
fact that the reconstruction procedure fixes only the spatially
dependent part of $\Phi_\pm$.

\paragraph{GLM reconstruction.}

At fixed time $t$, the scattering data consist of:

\begin{itemize}
\item the reflection coefficient $\mathscr{R}_\pm(t;\lambda)$ for
$\lambda \in \mathbb{R}$,
\item a discrete set of eigenvalues $\lambda_n^\pm = i\alpha_n^\pm$,
$\alpha_n^\pm>0$,
\item associated norming constants $\beta_n^\pm$.
\end{itemize}

From these data one constructs a kernel function $F_\pm(t,x)$ entering the
GLM equation
\begin{equation}
\label{eq:glm_equation_clean}
\begin{split}
     & K_\pm(t,x,y)
+
F_\pm(t,x+y)
\\
&+
\int_{\pm\infty}^{x}
K_\pm(t,x,z)\,
F_\pm(t,z+y)\,\dif z
=0 \, .
\end{split}
\end{equation}
Differentiating
\eqref{eq:glm_equation_clean} reproduces the differential constraint
\eqref{eq:kernel_pde_clean}, showing consistency of the formulation.

The inverse scattering transform is thus organized as the sequence
\begin{equation}
\{\mathscr{R}_\pm,\alpha_n^\pm,\beta_n^\pm\}
\longrightarrow
F_\pm
\longrightarrow
K_\pm
\longrightarrow
\mathcal{J}_\pm.
\end{equation}

Finally, the temporal part of the Lax pair induces a linear evolution of
the scattering data in the spectral parameter. This linearization is the
mechanism by which the nonlinear dynamics of $\mathcal{J}_\pm$ is mapped
to a simple flow in scattering space, characteristic of integrable
hierarchies.

\subsection{Single--soliton configuration}\label{sec:single-soliton-configuration}

We now consider the reflectionless sector of the inverse scattering problem,
characterized by a vanishing reflection coefficient, $\mathscr{R}_\pm(t,\lambda)=0$. In this case the scattering data are entirely determined by the discrete
spectrum of the associated Schr\"odinger operator together with the
corresponding norming constants. Since no continuous spectral contribution
is present, the GLM equation reduces to a
finite-rank integral equation.

For a single discrete eigenvalue in each chiral sector, let
\(\alpha_\pm>0\) denote the corresponding spectral parameter and
\(f_\pm(t)\) the associated phase function governing the time evolution.
The input kernel entering the GLM equation becomes separable and takes the
form
\begin{equation}
F_\pm(t,x)
=
\mp \,2\alpha_\pm\,
e^{
\mp\alpha_\pm (x-2f_\pm)}.
\end{equation}
The dependence on $x$ reflects the standard structure of the
Marchenko equation in the reflectionless sector. Owing to the separability
of $K_\pm$, the integral equation becomes algebraic and can be solved
exactly.

The corresponding Marchenko kernel is obtained as
\begin{equation}
K_\pm(t,x,y)
=
\pm
\frac{
2\alpha_\pm
e^{\pm\alpha_\pm(x-y)}
}
{
1+
e^{
\pm2\alpha_\pm\xi_\pm}
}
\,
\vartheta(\pm y\mp x),
\label{Marchenko_kernel}
\end{equation}
where $\xi_\pm = x-f_\pm(t)$. The Heaviside function imposes the support condition appropriate to the
GLM formulation and guarantees compatibility with the prescribed
asymptotic behavior at spatial infinity.

The physical field is reconstructed from the diagonal component of the
kernel through the standard reconstruction formula.
Substituting \eqref{Marchenko_kernel} yields the single--soliton profile
\begin{equation}
\mathcal{J}_\pm(\xi_\pm)
=
-\nu_\pm
\alpha_\pm^2
\,\mathrm{sech}^2
\bigl(
\alpha_\pm\xi_\pm
\bigr).
\label{eq:one-soliton}
\end{equation}
This solution is smooth and exponentially localized.
Its amplitude scales as \(\alpha_\pm^2\), while the characteristic width
is proportional to \(\alpha_\pm^{-1}\).
Increasing \(\alpha_\pm\) therefore produces a taller and narrower
configuration.

The reflectionless character of the solution implies the absence of
radiative modes. Consequently, the profile propagates without distortion:
its shape remains unchanged and only its center,
determined by \(f_\pm(t)\), evolves in time. This shape-preserving
property is the defining characteristic of a soliton and reflects the
integrable structure underlying the dynamics.

The associated spectral problem can be interpreted as a one-dimensional
Schr\"odinger equation. Since the two chiral sectors decouple, each sector gives rise to an
independent scattering problem.
For localized potentials such as
\eqref{eq:one-soliton}, the spectrum generally contains a continuous
component corresponding to scattering states together with a finite set of
bound states associated with the discrete eigenvalues.
In the present single--soliton configuration there exists a single discrete
bound state whose eigenvalue determines the soliton parameters.

It is instructive to relate these solutions to the bulk metric
\eqref{eq:ads3-metric-final}.
The single--soliton profile \eqref{eq:one-soliton} can be generated by an
appropriate choice of the boundary reparameterization functions
\(\Phi_\pm\),
\begin{equation}\label{O-S-CT}
\Phi_\pm(t,x)
=
\pm\frac{\pi \nu_\pm}{k}\,
\frac{
2\alpha_\pm
}{
1+
e^{\pm2\alpha_\pm\xi_\pm}
} + \varphi_\pm(t) \, .
\end{equation}
This construction shows that non-trivial solitonic configurations can be
generated through specific choices of boundary data while the bulk
spacetime remains locally AdS\(_3\).
The resulting geometries therefore do not correspond to localized bulk
sources; rather, they represent distinct configurations associated with
non-trivial boundary excitations.

\subsubsection{Spectral problem}

For the single--soliton configuration \eqref{eq:one-soliton}, the Schr\"odinger equation \eqref{Sch-eq} becomes exactly solvable. After a suitable rescaling of the coordinate $\xi_\pm$, it reduces to the P\"oschl--Teller potential with parameter $l=1$, a prototypical reflectionless system.

In this case, the spectral problem exhibits a simple structure: there is a single bound state associated with the localized soliton profile, while the continuous spectrum corresponds to purely transmitting scattering states.

\paragraph{Scattering States.}

For real $\lambda_\pm$, corresponding to $E_\pm>0$, equation \eqref{Sch-eq} admits two linearly independent scattering solutions. These can be obtained by dressing the free (Jost) solutions $\phi_\pm$, which asymptotically behave as plane waves, by a position--dependent factor induced by the potential.

Substituting \eqref{S-S-Free} and \eqref{Marchenko_kernel} into the Volterra representation \eqref{eq:jost_volterra_clean}, one obtains the explicit form
\begin{equation}\label{scattering-solution}
\psi_\pm=  \frac{\mathcal{C}_\pm(t;\lambda_\pm)\, e^{i\lambda_\pm x}}{ \lambda_\pm\pm i \alpha_\pm } \left[ \lambda_\pm + i\alpha_\pm\tanh\!\left(\alpha_\pm \xi_\pm\right)\right] ,
\end{equation}
together with its complex conjugate, which form a complete basis of solutions in the scattering sector.

In the asymptotic regions $x \to \pm\infty$, the hyperbolic tangent approaches constant values and the solutions reduce to plane waves, consistently matching the Jost behavior $\phi_\pm$. More precisely, the effect of the potential is to induce a phase shift without generating any reflected component.

\paragraph{Bound States.}

Bound states correspond to $E_\pm<0$ and arise when the spectral parameter $\lambda_\pm$ takes purely imaginary values. In this case, normalizability requires the solution to decay exponentially as $x \to \pm\infty$, which restricts the allowed values of $\lambda_\pm$ to a discrete set.

From the explicit form of the scattering solutions \eqref{scattering-solution}, one observes that at $\lambda_\pm=\pm i \alpha_\pm$ the normalization factor develops a simple pole. This signals the breakdown of the scattering description and indicates the emergence of a bound state. The corresponding discrete eigenvalue is $E_\pm=-\alpha_\pm^2/2$. Equivalently, this pole corresponds to a zero of the Jost function in the upper/lower half of the complex $\lambda_\pm$--plane, which is the standard criterion for the existence of bound states in one--dimensional scattering theory.

Extracting the regular part of the solution at the pole, one obtains the normalizable bound state wavefunction,
\begin{equation}
 \psi_\pm=\frac{1}{2}\,e^{\mp \alpha_\pm f_\pm (t)}\, \mathcal{C}_\pm(t,\pm i \alpha_\pm) \,  \text{sech}\!\left( \alpha_\pm \xi_\pm\right)\, .
\end{equation}
This solution is localized around $\xi_\pm=0$ and decays exponentially at spatial infinity.

\subsubsection{Fixing the time evolution}

The dynamics of the hierarchy is determined by the bi-Hamiltonian
structure \eqref{biHam}. Choosing the Hamiltonian associated with the
$(I+1)$-th flow fixes the chemical potentials to be
\begin{equation}
\mu_\pm=\mathcal{R}_{I+1}[\mathcal{J}_\pm]\, .
\end{equation}
Combining \eqref{eq:mu-to-Phi-clean} with \eqref{O-S-CT} then yields the
evolution equation
\begin{equation}
\mp \frac{\pi}{k}\,\mathcal{J}_\pm\,\partial_t f_\pm
\pm \partial_t\varphi_\pm
=
\mathcal{R}_{I+1}[\mathcal{J}_\pm] .
\label{eq:evolution_f_phi}
\end{equation}
For generic flows with $I\geq 0$, consistency of
\eqref{eq:evolution_f_phi} with the reconstruction formula
\eqref{reconstruction-formula} fixes the normalization constant to be
\begin{equation}
\nu_\pm=\frac{c_\pm}{6\pi}\, .
\end{equation}
The remaining freedom can be absorbed into a redefinition of the phase,
and therefore one may set $\varphi_\pm=0$,
without loss of generality.

Additional information follows from the anisotropic scaling symmetry
\eqref{eq:Lifshitz_scaling_refined}. Requiring both the Schr\"odinger
equation \eqref{Sch-eq} and the one-soliton configuration
\eqref{eq:one-soliton} to transform covariantly under the Lifshitz
rescaling uniquely determines the scaling weights of the parameters
appearing in the solution. One finds
\begin{equation}
\lambda_\pm\rightarrow \zeta\,\lambda_\pm,
\qquad
\alpha_\pm\rightarrow \zeta\,\alpha_\pm,
\qquad
f_\pm\rightarrow \zeta^{-1}f_\pm .
\end{equation}
Hence both the spectral parameter $\lambda_\pm$ and the inverse soliton
width $\alpha_\pm$ carry scaling dimension $+1$, while the position modulus $f_\pm$ has scaling dimension $-1$, as expected for a length
parameter.

Since $\alpha_\pm$ is constant along the flow, dimensional analysis
implies that the velocity of the soliton can depend only on the
combination $\alpha_\pm^{2I}$ characterizing the $(I+1)$-th flow.
Consequently the position modulus evolves linearly in time,
\begin{equation}
f_\pm(t)=\mp v_\pm t+x_\circ^\pm ,
\end{equation}
where $x_\circ^\pm$ is an integration constant. Substituting the one-soliton
solution into \eqref{eq:evolution_f_phi} determines the velocity
explicitly,
\begin{equation}
v_\pm^{I+1}=\frac{k}{\pi}
\left(
-\frac{c_\pm\alpha_\pm^2}{6\pi}
\right)^I .
\label{eq:soliton_velocity}
\end{equation}
Therefore each member of the hierarchy describes a rigidly propagating
soliton whose speed is fixed by the corresponding Hamiltonian flow.
The higher flows preserve the shape of the solution while modifying its
propagation velocity through a characteristic power of
$\alpha_\pm^2$.

\subsubsection{Hamiltonians}

The Lifshitz scaling symmetry also determines the functional dependence
of the conserved Hamiltonians on the soliton parameters. In the
single-soliton sector, the parameter $\alpha_\pm$ is the only
independent dimensionful quantity. Consequently, homogeneity under
\eqref{eq:Lifshitz_scaling_refined} requires the Hamiltonian generating
the $(I+1)$-th flow to scale as $H_{I+1}^\pm \propto \alpha_\pm^{\,2I+3}$.

This scaling behaviour admits a simple physical interpretation. Since
the velocity scales as $v_\pm\sim \alpha_\pm^{2I}$, the hierarchy
assigns increasingly higher scaling weight to the energy of a given
soliton as one moves to higher Hamiltonian flows. Thus the infinite set
of conserved charges may be viewed as different homogeneous measures of
the same solitonic excitation, each adapted to a particular Lifshitz
scaling exponent.

Evaluating the Hamiltonian densities on the one-soliton solution and
performing the spatial integration yields
\begin{equation}
H_{I}^\pm
=
\frac{2}{2I+1}
\left(
-\frac{c_\pm}{6\pi}
\right)^{I+1} \alpha_\pm^{2I+1}.
\label{eq:HI_soliton}
\end{equation}

Equation \eqref{eq:HI_soliton} explicitly realizes the scaling
prediction above and shows that all Hamiltonians of the hierarchy are
completely determined by a single parameter, namely the inverse width
$\alpha_\pm$ of the soliton.

\section{\texorpdfstring{$\sqrt{T\bar{T}}$}{TTbar} deformed integrable system}
\label{sec:ttbar-deformed-kdv}

In Secs.~\ref{sec:bi-hamiltonian-structure}--\ref{sec:lax-pair-formalism}, we established that the chiral currents $\mathcal{J}_\pm(t,x)$ satisfy an integrable Gel'fand--Dikii hierarchy governing their time evolution. Throughout that analysis, the radial coordinate plays no dynamical role: by construction, $\mathcal{J}_\pm$ are $r$-independent quantities parametrizing the radial-gauge connection~\eqref{eq:radialgauge}. On the other hand, Sec.~\ref{sec:ttbar} introduced an independent radial flow describing the evolution of the quasi-local observables $(\rho,J)$ between hypersurfaces of constant radius. Since both descriptions originate from the same bulk solution, it is natural to ask how the integrable hierarchy is perceived by an observer located at a finite radial cutoff. In particular, one may ask whether the finite-cutoff description induces a nontrivial deformation of the integrable dynamics or modifies it.

The first point to emphasize is that the Gel'fand--Dikii hierarchy itself is \emph{not} altered by the radial flow. As an evolution equation for the fundamental variables $\mathcal{J}_\pm(t,x)$, the hierarchy is completely independent of the choice of radial hypersurface. Consequently, every solution of the hierarchy, including its complete soliton sector, is identical to that obtained in the asymptotic theory. The radial evolution therefore does not generate a new integrable hierarchy, nor does it modify the dispersion relations or Hamiltonian flows governing $\mathcal{J}_\pm$. In this precise sense, there is no intrinsic {$\sqrt{T\bar{T}}$}-deformed Gel'fand--Dikii hierarchy'' acting on the chiral currents themselves.

The effect of the finite radial cutoff appears instead at the level of the quasi-local observables. Although the chiral currents remain the fundamental unconstrained degrees of freedom, the physical energy and momentum measured on a hypersurface at finite radius are nonlinear functionals of these currents. Consequently, the cutoff modifies the map between the integrable variables and the observable quantities, while leaving the underlying dynamics unchanged. Since the quasi-local stress tensor provides the natural holographic and fluid-dynamical observables discussed in Sec.~\ref{sec:fluid-gravity}, this is the appropriate sense in which the finite-cutoff theory realizes a {$\sqrt{T\bar{T}}$} deformation.

For the integrable system described in Secs.~\ref{sec:bi-hamiltonian-structure}--\ref{sec:lax-pair-formalism}, the quasi-local energy and momentum are given by
\begin{subequations}
\begin{align}
\mathscr{E}
&=
\frac{r^4+\ell^4}{r^4-\ell^4}
\, \mathscr{E}_{\infty}
+
\frac{4\ell^2 r^2}{r^4-\ell^4}
\, \mathscr{E}_{\text{int.}} \, ,
\label{QLEIS}
\\
\mathscr{J}
&=
-\frac{\pi}{k}
\left(
H_1^+-H_1^-
\right)\, .
\label{QLJIS}
\end{align}
\end{subequations}
where
\begin{subequations}
\begin{align}
\mathscr{E}_{\infty}
&=
\frac{\pi}{k}
\left(
H_1^++H_1^-
\right)\, ,
\\
\mathscr{E}_{\text{int.}}
&=
\frac{\pi}{2k}
\int_\Sigma
\dif x\,
\mathcal{J}_+\mathcal{J}_- \, .\label{E-int}
\end{align}
\end{subequations}

The quasi-local momentum retains a particularly simple form: it is given by the difference of the first Hamiltonians of the two chiral sectors and is therefore independent of the radial cutoff. The quasi-local energy exhibits a qualitatively different structure. Besides the contribution proportional to the total asymptotic energy $\mathscr{E}_\infty$, it contains an additional bilinear term involving both chiral currents. This interaction term is nonvanishing for every finite cutoff surface with $r>\ell$ and disappears only in the asymptotic limit $r\rightarrow\infty$, where the standard decomposition into independent left- and right-moving sectors is recovered.

Equation~\eqref{QLEIS} therefore demonstrates that the finite-cutoff theory does not couple the chiral sectors dynamically; rather, it couples them through the observable notion of energy. The currents $\mathcal{J}_+$ and $\mathcal{J}_-$ continue to evolve independently according to the same integrable hierarchy, preserving the complete bi-Hamiltonian structure and all associated conserved quantities. However, the quasi-local energy measured by an observer at finite radius is no longer the direct sum of the energies carried by the two sectors, but instead acquires an additional mixed contribution depending simultaneously on both chiral currents.

It is important to stress that this feature is exact and non-perturbative. Equation~\eqref{QLEIS} follows directly from the exact relation~\eqref{rho-final} and is valid for arbitrary values of the radial cutoff. The resulting coupling should therefore be understood as a purely quasi-local effect arising from the finite radial position of the observer, rather than as a modification of the underlying integrable dynamics. As the cutoff surface is moved towards the asymptotic boundary, the mixed contribution vanishes smoothly, and the conventional decomposition into two decoupled chiral sectors is recovered.

\subsection{Finite-cutoff soliton interaction}
\label{sec:two-soliton-interaction}

Equation~\eqref{E-int} attains its maximum when the centers of the two solitons coincide. As the separation between the solitons increases, the overlap between their localized profiles decreases exponentially, with the asymptotic decay governed by $\min(\alpha_+,\alpha_-)$. Consequently, the additional contribution to the quasi-local energy exhibits precisely the qualitative behavior expected of an interaction energy between two spatially localized objects, despite the absence of any direct coupling in the underlying equations of motion.

This interpretation becomes more transparent upon considering the total quasi-local energy $\mathscr{E}$. The contribution arising from the cross--term is
\begin{equation}\label{eq:interaction-energy}
    \begin{split}
        \mathscr{E}_{\rm int}
    =
    \frac{\pi}{2k}\, &\nu_+\nu_-  (\alpha_+\alpha_-)^2  \\
    & \times \int_{-\infty}^{+\infty}
    \left[\mathrm{sech}(\alpha_+\xi_+)\,\mathrm{sech}(\alpha_-\xi_-)\right]^2\,\dif x \, .
    \end{split}
\end{equation}

The remaining integral represents the overlap of two exponentially localized soliton profiles centered at $f_+(t)$ and $f_-(t)$, respectively. Since the integrand decays exponentially as $|x|\rightarrow\infty$, the integral is finite for arbitrary values of the separation. Introducing the instantaneous distance $\Delta(t)=|f_+-f_-|$,
one immediately observes that the overlap integral is a smooth function of $\Delta(t)$. It reaches its maximum when the two solitons coincide and decreases rapidly as their separation grows, becoming exponentially suppressed once $\Delta(t)$ exceeds the characteristic widths of the individual solitons, determined by $\alpha_\pm^{-1}$.

Equation~\eqref{eq:interaction-energy} therefore defines a finite, time-dependent contribution to the total quasi-local energy measured at a finite radial cutoff. As the two chiral solitons approach each other, the overlap of their profiles increases, generating an additional localized contribution to the energy. This contribution reaches its maximum when the solitons overlap and subsequently decreases as they separate. The effect originates entirely from the finite-cutoff contribution to the quasi-local stress tensor and has no counterpart in the asymptotic ($r\rightarrow\infty$) description.

This cutoff-induced energy should not be interpreted as evidence for a genuine dynamical interaction between the two solitons. The solitons propagate according to the undeformed dynamics, and the infinite families of conserved charges $H_I^\pm$ introduced in Sec.~\ref{sec:bi-hamiltonian-structure} remain separately conserved for every value of the cutoff. No additional force is generated, no energy is exchanged between the two chiral sectors, and the underlying integrable evolution is unaffected.

The role of the finite cutoff is therefore purely geometric: it modifies the relation between the local observables encoded in the quasi-local stress tensor and the physical energy measured on a finite-radius hypersurface. From this perspective, the quantity $\mathscr{E}_{\rm int}$ is an emergent interaction energy arising from the nonlinear mixing of the left- and right-moving sectors in the quasi-local energy functional, rather than from a modification of the microscopic equations governing the soliton dynamics. The resulting picture is particularly noteworthy: the bulk evolution remains exactly integrable, while finite-cutoff observables exhibit an effective, separation-dependent interaction that disappears continuously in the asymptotic limit.

\section{Discussion}\label{sec:Discussion}

\subsection{Spectral data as gravitational observables}

The inverse-scattering formulation developed in
Sec.~\ref{sec:lax-pair-formalism} identifies the chiral currents
$\mathcal{J}_\pm(t,x)$ with the potentials of an auxiliary
one-dimensional Schr\"odinger operator and parametrizes the classical
solution space in terms of the associated spectral data. Instead of
describing a configuration by the local boundary fields
$\mathcal{J}_\pm(t,x)$, one may equivalently specify the corresponding
reflection coefficient, the discrete bound-state eigenvalues, and the
associated norming constants in each chiral sector. Although this
reformulation originates from the mathematical theory of integrable
systems, every element of the spectral data admits a natural
gravitational interpretation. Moreover, it makes explicit why the
non-compact spatial topology, $\Sigma=\mathbb{R}$, plays a central role,
since the notion of scattering data requires asymptotic regions where
the auxiliary Schr\"odinger problem becomes free.

The chiral currents $\mathcal{J}_\pm(t,x)$ constitute the complete set
of gravitational boundary data. As shown in
Secs.~\ref{sec:three-dim-gravity}--\ref{sec:metric-formalism}, once
$\mathcal{J}_\pm(t,x)$ is specified, the bulk metric
\eqref{eq:ads3-metric-final} is uniquely determined up to the residual
gauge freedom encoded by $\Phi_\pm$. Consequently, the inverse-scattering
transform provides an alternative parametrization of the same
gravitational phase space,
\begin{equation}
    \{\mathcal{J}_\pm(t,x)\}
    \quad\longleftrightarrow\quad
    \{\mathscr{R}_\pm(t;\lambda),\ \alpha_n^\pm,\ \beta_n^\pm(t)\} \, .
\end{equation}
The GLM reconstruction described in
Sec.~\ref{sec:lax-pair-formalism} establishes that this correspondence
is one-to-one. Therefore, the spectral variables neither introduce new
degrees of freedom nor discard existing ones.

From the gravitational perspective, the spectral decomposition naturally
separates the boundary dynamics into two sectors associated with the
continuous and discrete spectra of the Schr\"odinger operator
\eqref{Sch-eq}.

\begin{itemize}

\item \emph{Continuous spectrum.}
The reflection coefficient
$\mathscr{R}_\pm(t;\lambda)$ characterizes the scattering states with
$\lambda_\pm^2\geq0$. Since the Schr\"odinger operator
$\mathrm{S}_\pm$ is constructed directly from the boundary currents
$\mathcal{J}_\pm$, which in turn determine the bulk metric through
\eqref{eq:ads3-metric-final}, the continuous spectrum encodes the
extended, nonlocalized sector of the boundary dynamics. These
configurations evolve under the integrable hierarchy without preserving
a localized profile. This interpretation concerns only boundary degrees
of freedom. Because three-dimensional Einstein gravity has no local
propagating bulk degrees of freedom, the continuous spectrum should not
be interpreted as describing bulk gravitons.

\item \emph{Discrete spectrum.}
The bound-state eigenvalues $\alpha_n^\pm$ together with the norming
constants $\beta_n^\pm(t)$ characterize the discrete spectrum
$\lambda_\pm^2<0$. These data describe spatially localized,
non-dispersive excitations of the boundary currents. A single bound
state reproduces the one-soliton solution discussed in
Sec.~\ref{sec:single-soliton-configuration}, while several discrete
eigenvalues generate multi-soliton configurations. The eigenvalues
determine the intrinsic parameters of the solitons, whereas the norming
constants specify their relative positions and phases and carry the
time dependence dictated by the isospectral evolution.

\end{itemize}

The inverse-scattering formulation therefore endows the gravitational
phase space with a distinguished decomposition into localized solitonic
configurations and extended scattering states. Under the isospectral
evolution generated by the Lax equation~\eqref{Lax-eq}, the discrete
eigenvalues remain invariant, while the reflection coefficient evolves
according to the corresponding hierarchy. The preservation of the
spectrum is the defining characteristic of integrability and underlies
the existence of the infinite family of commuting Hamiltonians
$H_I^\pm$ discussed in
Sec.~\ref{sec:bi-hamiltonian-structure}.

In the inverse-scattering theory of KdV-type hierarchies, the conserved
Hamiltonians may be expressed entirely in terms of the spectral data
through trace identities or, equivalently, through the asymptotic
expansion of the resolvent of the Schr\"odinger operator
$\mathrm{S}_\pm$
\cite{faddeev1987hamiltonian,dickey2003soliton}. Although the explicit
trace identities have not been derived in the conventions adopted here,
their existence implies that the commuting Hamiltonians $H_I^\pm$
admit an equivalent spectral representation. From this perspective, the
conservation of the Hamiltonians and the isospectral nature of the
Lax evolution are complementary manifestations of the same underlying
integrable structure.
Beyond their role in reconstructing the bulk geometry, the same spectral data also provide a natural framework for constructing integrable deformations of the boundary dynamics through self-consistent eigenfunction forcing, thereby extending the inverse-scattering description beyond the purely isospectral setting~ \cite{Adami:2025pfk}.

\subsection{Soliton solution}

The gravitational interpretation of the soliton solution
\eqref{eq:one-soliton} deserves particular attention. Throughout the
sector considered in this work, every solution is locally AdS$_3$, as
required by the bulk field equations $F=0$ discussed in
Sec.~\ref{sec:three-dim-gravity}. Consequently, the soliton does not
represent a new local bulk geometry produced by localized matter or
curvature. Rather, its physical content is encoded entirely in the
boundary data. More precisely, the profile
\eqref{eq:one-soliton} specifies a localized configuration of the
boundary currents $\mathcal{J}_\pm(t,x)$ entering the metric
\eqref{eq:ads3-metric-final}. The corresponding spacetime is therefore
obtained from global AdS$_3$ by a nontrivial large boundary
diffeomorphism parametrized by $\Phi_\pm$ through
\eqref{O-S-CT}. In this sense, the soliton belongs to the same class of
locally AdS$_3$ geometries discussed in
Sec.~\ref{sec:three-dim-gravity}; its distinguishing feature is the
presence of a spatially localized boundary excitation rather than a
homogeneous boundary configuration such as the BTZ solution.

The designation ``soliton'' follows from its dynamical properties.
As shown in
Sec.~\ref{sec:single-soliton-configuration}, the profile
$\mathrm{sech}^2(\alpha_\pm\xi_\pm)$ preserves its functional form
under the Hamiltonian evolution generated by every member of the
integrable hierarchy. Its amplitude and width remain constant, while
its center $f_\pm(t)$ evolves according to the velocity
\eqref{eq:soliton_velocity}. From the gravitational viewpoint, the
localized boundary excitation propagates rigidly along the boundary
without dispersion. Equivalently, the corresponding localized
deformation of the bulk metric determined by
\eqref{eq:ads3-metric-final} retains its shape throughout the
evolution. This is the gravitational realization of the familiar
balance between nonlinearity and dispersion in integrable systems,
encoded here in the nonlinear differential operator
$\mathrm{D}_\pm$ appearing in \eqref{Dpm-new}.

An important feature of the solution is that
$\mathcal{J}_\pm(\xi_\pm)\rightarrow0$ as
$\xi_\pm\rightarrow\pm\infty$, thereby satisfying the falloff
condition~\eqref{eq:finitecharges} required for a well-defined
non-compact phase space, while the zero-mode charge $H_0^\pm$ remains
finite and nonvanishing. This provides an explicit realization of the
general discussion in
Sec.~\ref{sec:killing-symmetries}. Since $H_0^\pm$ depends only on the
integrated zero mode of the boundary current, a sufficiently localized
configuration may carry finite conserved charges even though the fields
vanish asymptotically. From the holographic perspective developed in
Sec.~\ref{sec:fluid-gravity}, the soliton therefore represents a
localized excitation carrying a finite
$\hat{\mathfrak{u}}(1)$ charge or, equivalently, a finite value of the
Hamiltonian $H_0^\pm$ through \eqref{eq:HI_soliton} with $I=0$.

Equation~\eqref{eq:soliton_velocity} further shows that the spatial
profile of the soliton, determined solely by the parameter
$\alpha_\pm$, is universal throughout the hierarchy, whereas its
velocity depends on the Hamiltonian generating the evolution,
scaling as $v_\pm^{I+1}\propto\alpha_\pm^{2I}$. This reflects the
existence of infinitely many commuting Hamiltonian flows acting on the
same gravitational phase space. As discussed in
Sec.~\ref{sec:vanishing-symplectic-flux}, the Hamiltonian
$\mathcal{H}$ is fixed by the boundary conditions through
\eqref{H-Functional}. Different admissible boundary Hamiltonians
therefore generate distinct notions of boundary time evolution while
acting on the same family of geometries. Consequently, the same
localized gravitational excitation propagates with different velocities
under different members of the hierarchy, while preserving its shape
under each evolution.

\subsection{Non--compact boundaries}

As discussed in Sec.~\ref{sec:three-dim-gravity}, the spatial manifold
may be chosen either as the real line, $\Sigma=\mathbb{R}$, or as the
circle, $\Sigma=[0,L)$. This choice determines not only the global
structure of the gravitational phase space but also the appropriate
spectral description of the boundary dynamics. Consequently, it plays a
fundamental role in the construction developed in
Secs.~\ref{sec:bi-hamiltonian-structure}--\ref{sec:lax-pair-formalism}.

For compact spatial slices, the natural boundary condition is the
periodicity of the chiral currents $\mathcal{J}_\pm$. The
symmetry algebra derived in
Sec.~\ref{sec:symplectic-structure} is then naturally represented in
terms of Fourier modes and, after the inclusion of the central
extension, becomes the Virasoro algebra. Fourier analysis therefore
provides the appropriate description of the phase space containing the
BTZ black hole and its Virasoro descendants.

For the non-compact choice $\Sigma=\mathbb{R}$, periodicity is replaced
by falloff condition \eqref{eq:finitecharges}, requiring the
currents to decay sufficiently rapidly at spatial infinity. The
symmetry parameters $\epsilon_\pm(t,x)$ are consequently taken to be
smooth functions that vanish asymptotically. More importantly, the
spectral description changes qualitatively. The appropriate framework
is no longer Fourier analysis but inverse scattering theory (See also \cite{Latifi:2026ttmp}).

The Jost solutions, reflection and transmission coefficients, and the GLM
equation introduced in
Sec.~\ref{sec:lax-pair-formalism} all rely on the existence of
asymptotically free regions where the auxiliary Schr\"odinger equation
reduces to the free-particle problem. These structures are intrinsic to
the non-compact geometry and have no direct analogue for periodic
boundary conditions.

As emphasized in
Sec.~\ref{sec:killing-symmetries}, the zero-mode charges
$H_0^\pm$ distinguish disconnected sectors of the phase space and form
a continuous spectrum when the target space of the St\"uckelberg field
$\Phi_\pm$ is non-compact. The topology $\Sigma=\mathbb{R}$ naturally
realizes this situation. Since $\mathcal{J}_\pm$ is related to
$\Phi_\pm$ through \eqref{eq:J-to-Phi-clean}, the falloff condition
\eqref{eq:finitecharges} requires only that
$\mathcal{J}_\pm$ be integrable. Consequently,
$\Phi_\pm(+\infty)-\Phi_\pm(-\infty)$ is unrestricted and may assume
arbitrary real values compatible with the boundary conditions. The
soliton solutions therefore generate a continuous family of sectors
parametrized by the corresponding values of $H_0^\pm$. By contrast,
for a compact spatial circle, the periodicity of $\Phi_\pm$
generically restricts these sectors to discrete values, up to the
standard subtleties associated with winding sectors.

The inverse-scattering construction itself is intrinsically
linked to the non-compact geometry. The Jost solutions are defined by
their asymptotic plane-wave behavior,
$\phi_\pm(t,x;\lambda_\pm)\sim e^{i\lambda_\pm x}$ as
$x\rightarrow\pm\infty$, which requires the potential
$\mathcal{J}_\pm$ to vanish sufficiently rapidly so that the
Schr\"odinger problem \eqref{Sch-eq} becomes asymptotically free. No
such asymptotic regions exist on a spatial circle. Instead, the
appropriate spectral theory is formulated in terms of Floquet theory
and the periodic spectrum of Hill's equation \cite{Lazutkin1975NormalFA,Oblak:2016eij}. The corresponding inverse
problem gives rise to finite-gap and quasi-periodic solutions rather
than the bound-state spectrum underlying the soliton sector discussed
in Sec.~\ref{sec:lax-pair-formalism}. The non-compact topology is
therefore not merely a convenient technical choice but the natural
geometric setting in which inverse-scattering methods provide a
complete description of the gravitational phase space.

\subsection{Radial flow is not a Hamiltonian flow}
\label{sec:no-go-radial-hamiltonian}

The preceding discussion leads to an important conclusion: the radial evolution should not be interpreted as a Hamiltonian flow on
the same infinite-dimensional phase space that generates the commuting
$t$-flows. Indeed, in the radial gauge \eqref{eq:radialgauge}, the
chiral currents $\mathcal{J}_\pm(t,x)$ are, by construction,
independent of the radial coordinate. Consequently, the radial
coordinate does not generate a dynamical evolution of the phase space
coordinates $\mathcal{J}_\pm$; there is therefore no Hamiltonian vector
field associated with radial translations on this phase space.

Instead, the radial dependence enters only through the explicit,
pointwise algebraic relation between the chiral currents and
the physical observables measured on a finite-radius hypersurface.
Equivalently, for every fixed point of the phase space parametrized by
$\mathcal{J}_\pm$, the map
\begin{equation}
    (\mathcal{J}_+,\mathcal{J}_-)
\longmapsto
(\rho,J) \, ,
\end{equation}
depends explicitly on $r$, while the underlying phase space point
itself remains unchanged. From this perspective,
\eqref{eq:ttbar-flow-r} should be viewed not as a Hamiltonian evolution
equation but rather as an ordinary differential equation determining
the radial dependence of $\rho$ at fixed $J$ and fixed boundary coordinates $(t,x)$. Introducing the variable
$X:=\rho+\sqrt{\rho^2-J^2}$ reduces the equation to the linear form
\begin{equation}
    \partial_r \ln X=-\frac{4\ell^2r}{r^4-\ell^4},
\end{equation}
which is a homogeneous, $J$-independent first-order equation. Its
integration immediately reproduces the closed-form solution presented
in \eqref{rho-final}. The two integration constants specifying a
solution are precisely the momentum density $J$ and the asymptotic
energy density $\rho_\infty$ (or, equivalently, $X_\infty$). These
quantities label the underlying phase space point, while the radial
coordinate merely specifies how that fixed point is represented in
terms of the observables $(\rho,J)$ at a given cutoff surface.

An immediate consequence is that the complete integrable structure
constructed in Sec.~\ref{sec:bi-hamiltonian-structure} is unaffected by
the radial deformation. Since every conserved Hamiltonian
$H_I^\pm$ depends exclusively on the $r$-independent currents
$\mathcal{J}_\pm$, each member of the Gel'fand--Dikii hierarchy remains
exactly conserved for every value of the radial cutoff. No additional
$\sqrt{T\bar T}$ corrections arise in the infinite family of commuting
charges. Thus, the entire integrable hierarchy is common to all
finite-cutoff slices as well as to the asymptotic conformal boundary.
The only quantity that changes with the cutoff is the explicit
algebraic dictionary, given by
\eqref{rho-final}--\eqref{J-final}, relating the universal chiral data
to the locally measured energy density and momentum density.

This provides a precise characterization of the integrable structure
underlying the finite-cutoff theory. The radial {$\sqrt{T\bar{T}}$} deformation
does not generate a new Hamiltonian hierarchy, nor does it enlarge the
Gel'fand--Dikii bi-Hamiltonian structure by introducing an additional
commuting flow. Rather, the deformation is exactly solvable because the
radial evolution is purely algebraic at each spatial point, acting only
on the map between phase space variables and physical observables, while leaving the underlying integrable phase space
itself unchanged.

\section{Outlook}\label{sec:Outlook}

Several directions naturally follow from our analysis. A first step is to extend the present construction beyond the diagonal sector to the full phase space of three-dimensional gravity, where a richer boundary dynamics and symmetry algebra are expected. It would also be interesting to determine whether the spectral framework developed here applies to other integrable hierarchies induced by alternative boundary conditions, such as the Gardner, AKNS, and higher-spin hierarchies.
Related to this, the solitons are essentially classical objects, and the integrable hierarchy is generated by a Poisson structure, which should be ultimately deformed during a quantization procedure.  
It would be very interesting to investigate how quantum gravity effects, i.e. $1/c$ effects in 3D holography, enter during the quantization procedure and alter the integrable boundary structure. In particular the Dym hierarchy will be relevant in this regard, as perturbative deformations of the Brown--Henneaux boundary
conditions incorporating finite-$1/c$ corrections have been shown to give rise to this hierarchy in 
~\cite{Lara:2024cie}, revealing that integrability
persists even beyond the conventional semiclassical regime. Furthermore, quantum gravity effects in the bulk geometrize into island contributions,  \cite{Almheiri:2019qdq} related to the appearance of replica wormholes \cite{Penington:2019npb,Almheiri:2019psf,Penington:2019kki}. The appearance of islands requires a change to open boundary conditions, and it will be interesting to investigate which open boundary conditions will be compatible with the ones chosen for the appearance of the integrable structure.

Another interesting question for future research is whether  integrable hierarchies can appear in holography in dimensions higher than three, as well as in two dimensions. Many 2D dilaton gravities are known to appear as a dimensional reduction from higher dimensions \cite{grumiller2002dilaton,grumiller2006ramifications}. In particular, JT gravity arises as the leading two-dimensional effective theory obtained by dimensionally reducing the near-horizon region of a near-extremal BTZ black hole. More recently, the JT gravity dynamics has also been recovered through an off-shell decoupling limit without requiring dimensional reduction via a holographic renormalization group flow~\cite{Castro:2025itb}. While interesting progress has already been achieved through the realization of KdV structures in nearly AdS$_2$ gravity~\cite{Cardenas:2024hah}, investigating how further integrable structures descend to, or arise directly within, the two-dimensional setting hence is not only interesting in its own right, but also ties into the previous point on quantum gravity corrections, as island and replica wormhole contributions are most conveniently investigated in a \emph{2d} dilaton gravity setup such as JT gravity.

Moreover, as the integrability boundary conditions seem to remove most of the bulk dynamics, an interesting question is whether similar boundary conditions could simplify the situation in three-dimensional de Sitter holography \cite{Strominger:2001pn,Balasubramanian:2002zh,Anninos:2011ui,Chen:2022ozy}. Similarly, in flat space, one could ask what becomes of the integrability boundary conditions in AdS$_3$ after taking the flat space limit, and what their implications are for celestial holography in three dimensions \cite{Barnich:2006av,Barnich:2010eb,Bagchi:2014iea,Bagchi:2010zz,Bagchi:2012yk,Bagchi:2013lma,Barnich:2014zoa,Barnich:2014kra,Grumiller:2017sjh,Bagchi:2016bcd,Afshar:2019axx,Yu:2022bcp,Grumiller:2023rzn,Adami:2024rkr}.

An intriguing aspect of our results is the relation between the hierarchy of Hamiltonian charges and anisotropic time scalings. Each Hamiltonian defines an integrable flow with its own Lifshitz dynamical exponent, suggesting a hierarchy of effective time evolutions. Understanding the geometric and holographic origin of this correspondence remains an open problem. A related question is whether quantum deformations of the Poisson algebra of conserved  charges can lead to quantum corrections to the anisotropic scaling exponents.

The effective relativistic fluid description on a curved background may also admit connections with nonlinear fluid systems. Two-dimensional gravity is much less involved than its higher-dimensional counterparts, due to the absence of transverse shear. Restricting to conformal,\footnote{Without conformality, a bulk viscosity term is possible, but since this is $\propto \partial_\mu u^\mu$, a static flow needs pointlike sources of the velocity field, making it effectively non-smooth.} relativistic and uncharged fluids, there are no dissipative corrections at first in the derivative expansion at all, and this simple structure most probably continues to higher derivative order. These arguments however do not rule out the possibility of non-trivial non-dissipative transport coefficients such as the ones constructed using the generating functional methods of \cite{jensen2012towards,Banerjee:2012iz}.  In particular, these non-dissipative transport coefficients might be a deformation that do not spoil the non-dissipative behavior of soliton motion. It would hence be interesting to further investigate the relation between integrability and 1+1-dimensional hydrodynamics. This might help to better understand not only 3D holography, but also \emph{2d} hydrodynamics itself. Furthermore, a suitable formulation of this connection could provide effective descriptions of shallow-water waves, via analogue gravity \cite{Barcelo:2005fc}.

The inverse-scattering formulation further motivates a systematic study of the solution space. Beyond the single-soliton sector, it is important to construct general multi-soliton solutions and determine how finite-cutoff effects modify their interactions. Equally important is the continuous spectrum, whose gravitational interpretation and contribution to quasi-local observables remain largely unexplored. Random multi-soliton solutions might also be interesting from the point of view of one-dimensional turbulence, opening up a route to investigate, using methods similar to \cite{Dorband:2022bwv}, average dissipative behavior of such a fluid from the point of view of the AdS$_3$ bulk.

{Finally, our arguments show that the holographic radial flow is not generated by the canonical Hamiltonian structure of the boundary hierarchy. It is therefore natural to seek an alternative geometric formulation of the radial evolution, \emph{c.f.} \cite{Chen:2025jzb}.}

\section*{Acknowledgments}
We would like to thank H. Babaei-Aghbolagh for useful comments on the manuscript. K.L. gratefully acknowledges the warm hospitality of the Shanghai Institute for Mathematics and Interdisciplinary Sciences (SIMIS), where this project was conceived. R.M.~acknowledges the support of the German Research Foundation (DFG) through the Collaborative Research Center ToCoTronics, Project-ID 258499086 — SFB 1170, as well as Germany’s Excellence Strategy through the W{\"u}rzburg-Dresden Cluster of Excellence on Complexity and Topology in Quantum Matter - ctd.qmat (EXC 2147, Project-ID 390858490). 
R.M.~furthermore acknowledges hospitality of SIMIS and the associated travel support under STCSM Grant 25HB2701900.


\bibliographystyle{fullsort.bst}
\bibliography{reference}

@article{jensen2012towards,
  title={Towards hydrodynamics without an entropy current},
  author={Jensen, Kristan and Kaminski, Matthias and Kovtun, Pavel and Meyer, Rene and Ritz, Adam and Yarom, Amos},
  journal={Physical review letters},
  volume={109},
  number={10},
  pages={101601},
  year={2012},
  publisher={APS}
}

@article{grumiller2006ramifications,
  title={Ramifications of lineland},
  author={Grumiller, Daniel and Meyer, Rene},
  journal={Turkish Journal of Physics},
  volume={30},
  number={5},
  pages={349--378},
  year={2006}
}

@article{grumiller2002dilaton,
  title={Dilaton gravity in two dimensions},
  author={Grumiller, Daniel and Kummer, Wolfgang and Vassilevich, DV},
  journal={Physics Reports},
  volume={369},
  number={4},
  pages={327--430},
  year={2002},
  publisher={Elsevier}
}

@article{Dorband:2022bwv,
    author = "Dorband, Moritz and Grumiller, Daniel and Meyer, Ren{\'e} and Zhao, Suting",
    title = "{Disorder in AdS$_3$/CFT$_2$}",
    eprint = "2204.00596",
    archivePrefix = "arXiv",
    primaryClass = "hep-th",
    reportNumber = "TUW-22-04",
    doi = "10.21468/SciPostPhys.16.1.017",
    journal = "SciPost Phys.",
    volume = "16",
    number = "1",
    pages = "017",
    year = "2024"
}

@article{Banerjee:2012iz,
    author = "Banerjee, Nabamita and Bhattacharya, Jyotirmoy and Bhattacharyya, Sayantani and Jain, Sachin and Minwalla, Shiraz and Sharma, Tarun",
    title = "{Constraints on Fluid Dynamics from Equilibrium Partition Functions}",
    eprint = "1203.3544",
    archivePrefix = "arXiv",
    primaryClass = "hep-th",
    reportNumber = "TFR-TH-12-05, IPMU12-0037",
    doi = "10.1007/JHEP09(2012)046",
    journal = "JHEP",
    volume = "09",
    pages = "046",
    year = "2012"
}

@article{Henneaux:1985tv,
    author = "Henneaux, M. and Teitelboim, C.",
    title = "{Asymptotically anti-De Sitter Spaces}",
    doi = "10.1007/BF01205790",
    journal = "Commun. Math. Phys.",
    volume = "98",
    pages = "391--424",
    year = "1985"
}

@article{Banados:1993ur,
    author = "Banados, Maximo and Teitelboim, Claudio and Zanelli, Jorge",
    title = "{Dimensionally continued black holes}",
    eprint = "gr-qc/9307033",
    archivePrefix = "arXiv",
    reportNumber = "IASSNS-AST-93-45",
    doi = "10.1103/PhysRevD.49.975",
    journal = "Phys. Rev. D",
    volume = "49",
    pages = "975--986",
    year = "1994"
}

@article{Banados:1992wn,
    author = "Banados, Maximo and Teitelboim, Claudio and Zanelli, Jorge",
    title = "{The Black hole in three-dimensional space-time}",
    eprint = "hep-th/9204099",
    archivePrefix = "arXiv",
    reportNumber = "PRINT-92-0151 (CHILE), IASSNS-HEP-92-29",
    doi = "10.1103/PhysRevLett.69.1849",
    journal = "Phys. Rev. Lett.",
    volume = "69",
    pages = "1849--1851",
    year = "1992"
}

@book{Carlip:1998uc,
    author = "Carlip, Steven",
    title = "{Quantum gravity in 2+1 dimensions}",
    doi = "10.1017/CBO9780511564192",
    isbn = "978-0-521-54588-4, 978-0-511-82229-2",
    publisher = "Cambridge University Press",
    series = "Cambridge Monographs on Mathematical Physics",
    month = "12",
    year = "2003"
}

@article{Rodriguez:2021tcz,
    author = "Rodr{\'\i}guez, Pablo and Tempo, David and Troncoso, Ricardo",
    title = "{Mapping relativistic to ultra/non-relativistic conformal symmetries in 2D and finite $ \sqrt{T\overline{T}} $ deformations}",
    eprint = "2106.09750",
    archivePrefix = "arXiv",
    primaryClass = "hep-th",
    reportNumber = "CECS-PHY-20/03",
    doi = "10.1007/JHEP11(2021)133",
    journal = "JHEP",
    volume = "11",
    pages = "133",
    year = "2021"
}

@article{Banerjee:2026qyc,
    author = "Banerjee, Aritra and Parekh, Pulastya and Raj, Robin",
    title = "{On $ \sqrt{T\overline{T}} $ deformed pathways: CFT to CCFT}",
    eprint = "2601.15376",
    archivePrefix = "arXiv",
    primaryClass = "hep-th",
    doi = "10.1007/JHEP05(2026)267",
    journal = "JHEP",
    volume = "05",
    pages = "267",
    year = "2026"
}

@article{Ebert:2023tih,
    author = "Ebert, Stephen and Ferko, Christian and Sun, Zhengdi",
    title = "{Root-TT{\textasciimacron} deformed boundary conditions in holography}",
    eprint = "2304.08723",
    archivePrefix = "arXiv",
    primaryClass = "hep-th",
    doi = "10.1103/PhysRevD.107.126022",
    journal = "Phys. Rev. D",
    volume = "107",
    number = "12",
    pages = "126022",
    year = "2023"
}

@article{Tempo:2022ndz,
    author = "Tempo, David and Troncoso, Ricardo",
    title = "{Nonlinear automorphism of the conformal algebra in 2D and continuous $ \sqrt{T\overline{T}} $ deformations}",
    eprint = "2210.00059",
    archivePrefix = "arXiv",
    primaryClass = "hep-th",
    reportNumber = "CECS-PHY-22/05",
    doi = "10.1007/JHEP12(2022)129",
    journal = "JHEP",
    volume = "12",
    pages = "129",
    year = "2022"
}

@article{Balasubramanian:2002zh,
    author = "Balasubramanian, Vijay and de Boer, Jan and Minic, Djordje",
    editor = "de Wit, B. and Vandoren, S.",
    title = "{Notes on de Sitter space and holography}",
    eprint = "hep-th/0207245",
    archivePrefix = "arXiv",
    reportNumber = "VPI-IPPAP-02-05, UPR-1008-T, IFTA-2002-26",
    doi = "10.1016/S0003-4916(02)00020-9",
    journal = "Class. Quant. Grav.",
    volume = "19",
    pages = "5655--5700",
    year = "2002"
}

@article{Anninos:2011ui,
    author = "Anninos, Dionysios and Hartman, Thomas and Strominger, Andrew",
    title = "{Higher Spin Realization of the dS/CFT Correspondence}",
    eprint = "1108.5735",
    archivePrefix = "arXiv",
    primaryClass = "hep-th",
    doi = "10.1088/1361-6382/34/1/015009",
    journal = "Class. Quant. Grav.",
    volume = "34",
    number = "1",
    pages = "015009",
    year = "2017"
}

@article{Chen:2022ozy,
    author = "Chen, Heng-Yu and Hikida, Yasuaki",
    title = "{Three-Dimensional de Sitter Holography and Bulk Correlators at Late Time}",
    eprint = "2204.04871",
    archivePrefix = "arXiv",
    primaryClass = "hep-th",
    reportNumber = "YITP-22-37",
    doi = "10.1103/PhysRevLett.129.061601",
    journal = "Phys. Rev. Lett.",
    volume = "129",
    number = "6",
    pages = "061601",
    year = "2022"
}

@article{Strominger:2001pn,
    author = "Strominger, Andrew",
    title = "{The dS / CFT correspondence}",
    eprint = "hep-th/0106113",
    archivePrefix = "arXiv",
    doi = "10.1088/1126-6708/2001/10/034",
    journal = "JHEP",
    volume = "10",
    pages = "034",
    year = "2001"
}

@article{Conti:2022egv,
    author = "Conti, Riccardo and Romano, Jacopo and Tateo, Roberto",
    title = "{Metric approach to a $ \mathrm{T}\overline{\mathrm{T}} $-like deformation in arbitrary dimensions}",
    eprint = "2206.03415",
    archivePrefix = "arXiv",
    primaryClass = "hep-th",
    doi = "10.1007/JHEP09(2022)085",
    journal = "JHEP",
    volume = "09",
    pages = "085",
    year = "2022"
}

@article{Babaei-Aghbolagh:2022leo,
    author = "Babaei-Aghbolagh, H. and Babaei Velni, Komeil and Mahdavian Yekta, Davood and Mohammadzadeh, Hosein",
    title = "{Marginal TT{\textasciimacron}-like deformation and modified Maxwell theories in two dimensions}",
    eprint = "2206.12677",
    archivePrefix = "arXiv",
    primaryClass = "hep-th",
    doi = "10.1103/PhysRevD.106.086022",
    journal = "Phys. Rev. D",
    volume = "106",
    number = "8",
    pages = "086022",
    year = "2022"
}

@article{Babaei-Aghbolagh:2022uij,
    author = "Babaei-Aghbolagh, H. and Velni, Komeil Babaei and Yekta, Davood Mahdavian and Mohammadzadeh, H.",
    title = "{Emergence of non-linear electrodynamic theories from TT{\textasciimacron}-like deformations}",
    eprint = "2202.11156",
    archivePrefix = "arXiv",
    primaryClass = "hep-th",
    reportNumber = "IPM/P-2022/13",
    doi = "10.1016/j.physletb.2022.137079",
    journal = "Phys. Lett. B",
    volume = "829",
    pages = "137079",
    year = "2022"
}

@article{Lazutkin1975NormalFA,
  title={Normal forms and versal deformations for Hill's equation},
  author={Vladimir F. Lazutkin and T. F. Pankratova},
  journal={Functional Analysis and Its Applications},
  year={1975},
  volume={9},
  pages={306-311},
  url={https://api.semanticscholar.org/CorpusID:122001930}
}

@article{Gonzalez:2018jgp,
    author = "Gonz{\'a}lez, Hern{\'a}n A. and Matulich, Javier and Pino, Miguel and Troncoso, Ricardo",
    title = "{Revisiting the asymptotic dynamics of General Relativity on AdS$_{3}$}",
    eprint = "1809.02749",
    archivePrefix = "arXiv",
    primaryClass = "hep-th",
    reportNumber = "CECS-PHY-18/03",
    doi = "10.1007/JHEP12(2018)115",
    journal = "JHEP",
    volume = "12",
    pages = "115",
    year = "2018"
}

@article{Ojeda:2019xih,
    author = "Ojeda, Emilio and P{\'e}rez, Alfredo",
    title = "{Boundary conditions for General Relativity in three-dimensional spacetimes, integrable systems and the \text{KdV}/m\text{KdV} hierarchies}",
    eprint = "1906.11226",
    archivePrefix = "arXiv",
    primaryClass = "hep-th",
    reportNumber = "CECS-PHY-19/02",
    doi = "10.1007/JHEP08(2019)079",
    journal = "JHEP",
    volume = "08",
    pages = "079",
    year = "2019"
}

@article{Ferko:2022cix,
    author = "Ferko, Christian and Sfondrini, Alessandro and Smith, Liam and Tartaglino-Mazzucchelli, Gabriele",
    title = "{Root-$T \bar T$ Deformations in Two-Dimensional Quantum Field Theories}",
    eprint = "2206.10515",
    archivePrefix = "arXiv",
    primaryClass = "hep-th",
    doi = "10.1103/PhysRevLett.129.201604",
    journal = "Phys. Rev. Lett.",
    volume = "129",
    number = "20",
    pages = "201604",
    year = "2022"
}

@article{Fuentealba:2017omf,
    author = "Fuentealba, Oscar and Matulich, Javier and P{\'e}rez, Alfredo and Pino, Miguel and Rodr{\'\i}guez, Pablo and Tempo, David and Troncoso, Ricardo",
    title = "{Integrable systems with BMS$_{3}$ Poisson structure and the dynamics of locally flat spacetimes}",
    eprint = "1711.02646",
    archivePrefix = "arXiv",
    primaryClass = "hep-th",
    reportNumber = "CECS-PHY-17-02",
    doi = "10.1007/JHEP01(2018)148",
    journal = "JHEP",
    volume = "01",
    pages = "148",
    year = "2018"
}

@article{Afshar:2019axx,
    author = "Afshar, Hamid and Gonz{\'a}lez, Hern{\'a}n A. and Grumiller, Daniel and Vassilevich, Dmitri",
    title = "{Flat space holography and the complex Sachdev-Ye-Kitaev model}",
    eprint = "1911.05739",
    archivePrefix = "arXiv",
    primaryClass = "hep-th",
    reportNumber = "TUW-19-04",
    doi = "10.1103/PhysRevD.101.086024",
    journal = "Phys. Rev. D",
    volume = "101",
    number = "8",
    pages = "086024",
    year = "2020"
}

@article{Barnich:2014zoa,
    author = "Barnich, Glenn and Oblak, Blagoje",
    title = "{Holographic positive energy theorems in three-dimensional gravity}",
    eprint = "1403.3835",
    archivePrefix = "arXiv",
    primaryClass = "hep-th",
    doi = "10.1088/0264-9381/31/15/152001",
    journal = "Class. Quant. Grav.",
    volume = "31",
    pages = "152001",
    year = "2014"
}

@article{Grumiller:2023rzn,
    author = "Grumiller, Daniel and Riegler, Max",
    title = "{Carrollian c functions and flat space holographic RG flows in BMS3/CCFT2}",
    eprint = "2309.11539",
    archivePrefix = "arXiv",
    primaryClass = "hep-th",
    reportNumber = "TUW-23-04",
    doi = "10.1103/PhysRevD.108.126008",
    journal = "Phys. Rev. D",
    volume = "108",
    number = "12",
    pages = "126008",
    year = "2023"
}

@article{Bagchi:2016bcd,
    author = "Bagchi, Arjun and Basu, Rudranil and Kakkar, Ashish and Mehra, Aditya",
    title = "{Flat Holography: Aspects of the dual field theory}",
    eprint = "1609.06203",
    archivePrefix = "arXiv",
    primaryClass = "hep-th",
    doi = "10.1007/JHEP12(2016)147",
    journal = "JHEP",
    volume = "12",
    pages = "147",
    year = "2016"
}

@article{Adami:2024rkr,
    author = "Adami, H. and Sheikh-Jabbari, M. M. and Taghiloo, V.",
    title = "{Gravitational stress tensor and current at null infinity in three dimensions}",
    eprint = "2405.00149",
    archivePrefix = "arXiv",
    primaryClass = "hep-th",
    doi = "10.1016/j.physletb.2024.138835",
    journal = "Phys. Lett. B",
    volume = "855",
    pages = "138835",
    year = "2024"
}

@book{faddeev1987hamiltonian,
  title={Hamiltonian methods in the theory of solitons},
  author={Faddeev, Ludwig D and Takhtajan, Leon A},
  volume={23},
  year={1987},
  publisher={Springer}
}

@article{Grumiller:2019tyl,
    author = "Grumiller, Daniel and Merbis, Wout",
    title = "{Near horizon dynamics of three dimensional black holes}",
    eprint = "1906.10694",
    archivePrefix = "arXiv",
    primaryClass = "hep-th",
    reportNumber = "TUW-19-01",
    doi = "10.21468/SciPostPhys.8.1.010",
    journal = "SciPost Phys.",
    volume = "8",
    number = "1",
    pages = "010",
    year = "2020"
}

@article{Babaei-Aghbolagh:2024hti,
    author = "Babaei-Aghbolagh, H. and He, Song and Morone, Tommaso and Ouyang, Hao and Tateo, Roberto",
    title = "{Geometric Formulation of Generalized Root-TT{\textasciimacron} Deformations}",
    eprint = "2405.03465",
    archivePrefix = "arXiv",
    primaryClass = "hep-th",
    doi = "10.1103/PhysRevLett.133.111602",
    journal = "Phys. Rev. Lett.",
    volume = "133",
    number = "11",
    pages = "111602",
    year = "2024"
}

@article{Bhattacharyya:2008xc,
    author = "Bhattacharyya, Sayantani and Hubeny, Veronika E. and Loganayagam, R. and Mandal, Gautam and Minwalla, Shiraz and Morita, Takeshi and Rangamani, Mukund and Reall, Harvey S.",
    title = "{Local Fluid Dynamical Entropy from Gravity}",
    eprint = "0803.2526",
    archivePrefix = "arXiv",
    primaryClass = "hep-th",
    doi = "10.1088/1126-6708/2008/06/055",
    journal = "JHEP",
    volume = "06",
    pages = "055",
    year = "2008"
}

@article{Lara:2024cie,
    author = "Lara, Kristiansen and Pino, Miguel and Reyes, Francisco",
    title = "{1/c deformations of AdS$_{3}$ boundary conditions and the Dym hierarchy}",
    eprint = "2401.12338",
    archivePrefix = "arXiv",
    primaryClass = "hep-th",
    doi = "10.1007/JHEP11(2024)042",
    journal = "JHEP",
    volume = "11",
    pages = "042",
    year = "2024"
}

@article{Dymarsky:2020tjh,
    author = "Dymarsky, Anatoly and Sugishita, Sotaro",
    title = "{KdV-charged black holes}",
    eprint = "2002.08368",
    archivePrefix = "arXiv",
    primaryClass = "hep-th",
    doi = "10.1007/JHEP05(2020)041",
    journal = "JHEP",
    volume = "05",
    pages = "041",
    year = "2020"
}

@article{Arenas-Henriquez:2024ypo,
    author = "Arenas-Henriquez, Gabriel and Diaz, Felipe and Rivera-Betancour, David",
    title = "{Generalized Fefferman-Graham gauge and boundary Weyl structures}",
    eprint = "2411.12513",
    archivePrefix = "arXiv",
    primaryClass = "hep-th",
    doi = "10.1007/JHEP02(2025)007",
    journal = "JHEP",
    volume = "02",
    pages = "007",
    year = "2025"
}

@article{Cardenas:2021vwo,
    author = "C{\'a}rdenas, Marcela and Correa, Francisco and Lara, Kristiansen and Pino, Miguel",
    title = "{Integrable Systems and Spacetime Dynamics}",
    eprint = "2104.09676",
    archivePrefix = "arXiv",
    primaryClass = "hep-th",
    doi = "10.1103/PhysRevLett.127.161601",
    journal = "Phys. Rev. Lett.",
    volume = "127",
    number = "16",
    pages = "161601",
    year = "2021"
}

@book{ablowitz1981solitons,
  title={Solitons and the inverse scattering transform},
  author={Ablowitz, Mark J and Segur, Harvey},
  year={1981},
  publisher={SIAM}
}

@book{olver1993applications,
  title={Applications of Lie groups to differential equations},
  author={Olver, Peter J},
  volume={107},
  year={1993},
  publisher={Springer Science \& Business Media}
}

@article{lax1968integrals,
  title={Integrals of nonlinear equations of evolution and solitary waves},
  author={Lax, Peter D},
  journal={Communications on pure and applied mathematics},
  volume={21},
  number={5},
  pages={467--490},
  year={1968},
  publisher={Wiley Online Library}
}

@article{Rangamani:2009xk,
    author = "Rangamani, Mukund",
    editor = "Uranga, A. M.",
    title = "{Gravity and Hydrodynamics: Lectures on the fluid-gravity correspondence}",
    eprint = "0905.4352",
    archivePrefix = "arXiv",
    primaryClass = "hep-th",
    reportNumber = "NSF-KITP-09-65",
    doi = "10.1088/0264-9381/26/22/224003",
    journal = "Class. Quant. Grav.",
    volume = "26",
    pages = "224003",
    year = "2009"
}

@article{Zamolodchikov:2004ce,
    author = "Zamolodchikov, Alexander B.",
    title = "{Expectation value of composite field T anti-T in two-dimensional quantum field theory}",
    eprint = "hep-th/0401146",
    archivePrefix = "arXiv",
    reportNumber = "BONN-TH-2004-02",
    month = "1",
    year = "2004"
}

@article{Smirnov:2016lqw,
    author = "Smirnov, F. A. and Zamolodchikov, A. B.",
    title = "{On space of integrable quantum field theories}",
    eprint = "1608.05499",
    archivePrefix = "arXiv",
    primaryClass = "hep-th",
    doi = "10.1016/j.nuclphysb.2016.12.014",
    journal = "Nucl. Phys. B",
    volume = "915",
    pages = "363--383",
    year = "2017"
}

@article{Barnich:2001jy,
      author         = "Barnich, Glenn and Brandt, Friedemann",
      title          = "{Covariant theory of asymptotic symmetries, conservation
                        laws and central charges}",
      journal        = "Nucl. Phys.",
      volume         = "B633",
      year           = "2002",
      pages          = "3-82",
      doi            = "10.1016/S0550-3213(02)00251-1",
      eprint         = "hep-th/0111246",
      archivePrefix  = "arXiv",
      primaryClass   = "hep-th",
      reportNumber   = "ULB-TH-01-19, MPI-MIS-94-2001",
      SLACcitation   = "%%CITATION = HEP-TH/0111246;%%"
}

@article{Afshar:2016uax,
      author         = "Afshar, H. and Grumiller, D. and Sheikh-Jabbari, M. M.",
      title          = "{Near horizon soft hair as microstates of three
                        dimensional black holes}",
      journal        = "Phys. Rev.",
      volume         = "D96",
      year           = "2017",
      number         = "8",
      pages          = "084032",
      doi            = "10.1103/PhysRevD.96.084032",
      eprint         = "1607.00009",
      archivePrefix  = "arXiv",
      primaryClass   = "hep-th",
      reportNumber   = "TUW-16-12",
      SLACcitation   = "%%CITATION = ARXIV:1607.00009;%%"
}

@article{Perez:2016vqo,
      author         = "P{\'e}rez, Alfredo and Tempo, David and Troncoso, Ricardo",
      title          = "{Boundary conditions for General Relativity on AdS$_{3}$
                        and the \text{KdV} hierarchy}",
      journal        = "JHEP",
      volume         = "06",
      year           = "2016",
      pages          = "103",
      doi            = "10.1007/JHEP06(2016)103",
      eprint         = "1605.04490",
      archivePrefix  = "arXiv",
      primaryClass   = "hep-th",
      reportNumber   = "CECS-PHY-16-04",
      SLACcitation   = "%%CITATION = ARXIV:1605.04490;%%"
}

@article{Papadimitriou:2004ap,
      author         = "Papadimitriou, Ioannis and Skenderis, Kostas",
      title          = "{AdS / CFT correspondence and geometry}",
      booktitle      = "{AdS/CFT correspondence: Einstein metrics and their
                        conformal boundaries. Proceedings, 73rd Meeting of
                        Theoretical Physicists and Mathematicians, Strasbourg,
                        France, September 11-13, 2003}",
      journal        = "IRMA Lect. Math. Theor. Phys.",
      volume         = "8",
      year           = "2005",
      pages          = "73-101",
      doi            = "10.4171/013-1/4",
      eprint         = "hep-th/0404176",
      archivePrefix  = "arXiv",
      primaryClass   = "hep-th",
      reportNumber   = "ITFA-2004-17",
      SLACcitation   = "%%CITATION = HEP-TH/0404176;%%"
}

@article{Elitzur:1989nr,
      author         = "Elitzur, Shmuel and Moore, Gregory W. and Schwimmer, Adam
                        and Seiberg, Nathan",
      title          = "{Remarks on the Canonical Quantization of the
                        Chern-Simons-Witten Theory}",
      journal        = "Nucl. Phys.",
      volume         = "B326",
      year           = "1989",
      pages          = "108-134",
      doi            = "10.1016/0550-3213(89)90436-7",
      reportNumber   = "IASSNS-HEP-89/20",
      SLACcitation   = "%%CITATION = NUPHA,B326,108;%%"
}

@article{Afshar:2016wfy,
      author         = "Afshar, Hamid and Detournay, Stephane and Grumiller,
                        Daniel and Merbis, Wout and Perez, Alfredo and Tempo,
                        David and Troncoso, Ricardo",
      title          = "{Soft Heisenberg hair on black holes in three
                        dimensions}",
      journal        = "Phys. Rev.",
      volume         = "D93",
      year           = "2016",
      number         = "10",
      pages          = "101503",
      doi            = "10.1103/PhysRevD.93.101503",
      eprint         = "1603.04824",
      archivePrefix  = "arXiv",
      primaryClass   = "hep-th",
      reportNumber   = "TUW-16-06",
      SLACcitation   = "%%CITATION = ARXIV:1603.04824;%%"
}

@article{Bagchi:2014iea,
      author         = "Bagchi, Arjun and Basu, Rudranil and Grumiller, Daniel
                        and Riegler, Max",
      title          = "{Entanglement entropy in Galilean conformal field
                        theories and flat holography}",
      journal        = "Phys.Rev.Lett.",
      number         = "11",
      volume         = "114",
      pages          = "111602",
      doi            = "10.1103/PhysRevLett.114.111602",
      year           = "2015",
      eprint         = "1410.4089",
      archivePrefix  = "arXiv",
      primaryClass   = "hep-th",
      reportNumber   = "TUW-14-14",
      SLACcitation   = "%%CITATION = ARXIV:1410.4089;%%",
}

@article{Barnich:2014kra,
      author         = "Barnich, Glenn and Oblak, Blagoje",
      title          = "{Notes on the BMS group in three dimensions: I. Induced
                        representations}",
      journal        = "JHEP",
      volume         = "1406",
      pages          = "129",
      doi            = "10.1007/JHEP06(2014)129",
      year           = "2014",
      eprint         = "1403.5803",
      archivePrefix  = "arXiv",
      primaryClass   = "hep-th",
      SLACcitation   = "%%CITATION = ARXIV:1403.5803;%%",
}

@phdthesis{Oblak:2016eij,
    author = "Oblak, Blagoje",
    title = "{BMS Particles in Three Dimensions}",
    eprint = "1610.08526",
    archivePrefix = "arXiv",
    primaryClass = "hep-th",
    doi = "10.1007/978-3-319-61878-4",
    school = "Brussels U.",
    year = "2016"
}

@article{Barnich:2010eb,
      author         = "Barnich, Glenn and Troessaert, Cedric",
      title          = "{Aspects of the BMS/CFT correspondence}",
      journal        = "JHEP",
      volume         = "1005",
      pages          = "062",
      doi            = "10.1007/JHEP05(2010)062",
      year           = "2010",
      eprint         = "1001.1541",
      archivePrefix  = "arXiv",
      primaryClass   = "hep-th",
      reportNumber   = "ULB-TH-09-28",
      SLACcitation   = "%%CITATION = ARXIV:1001.1541;%%",
}

@inproceedings{Hubeny:2011hd,
    author = "Hubeny, Veronika E. and Minwalla, Shiraz and Rangamani, Mukund",
    title = "{The fluid/gravity correspondence}",
    booktitle = "{Theoretical Advanced Study Institute in Elementary Particle Physics}: {String theory and its Applications: From meV to the Planck Scale}",
    eprint = "1107.5780",
    archivePrefix = "arXiv",
    primaryClass = "hep-th",
    pages = "348--383",
    year = "2012"
}

@article{Bhattacharyya:2008ji,
    author = "Bhattacharyya, Sayantani and Loganayagam, R. and Minwalla, Shiraz and Nampuri, Suresh and Trivedi, Sandip P. and Wadia, Spenta R.",
    title = "{Forced Fluid Dynamics from Gravity}",
    eprint = "0806.0006",
    archivePrefix = "arXiv",
    primaryClass = "hep-th",
    doi = "10.1088/1126-6708/2009/02/018",
    journal = "JHEP",
    volume = "02",
    pages = "018",
    year = "2009"
}

@article{Erdmenger:2008rm,
    author = "Erdmenger, Johanna and Haack, Michael and Kaminski, Matthias and Yarom, Amos",
    title = "{Fluid dynamics of R-charged black holes}",
    eprint = "0809.2488",
    archivePrefix = "arXiv",
    primaryClass = "hep-th",
    reportNumber = "LMU-ASC-48-08, MPP-2008-116",
    doi = "10.1088/1126-6708/2009/01/055",
    journal = "JHEP",
    volume = "01",
    pages = "055",
    year = "2009"
}

@article{Banerjee:2008th,
    author = "Banerjee, Nabamita and Bhattacharya, Jyotirmoy and Bhattacharyya, Sayantani and Dutta, Suvankar and Loganayagam, R. and Surowka, P.",
    title = "{Hydrodynamics from charged black branes}",
    eprint = "0809.2596",
    archivePrefix = "arXiv",
    primaryClass = "hep-th",
    doi = "10.1007/JHEP01(2011)094",
    journal = "JHEP",
    volume = "01",
    pages = "094",
    year = "2011"
}

@article{Wald:1999wa,
      author         = "Wald, Robert M. and Zoupas, Andreas",
      title          = "{A General definition of 'conserved quantities' in
                        general relativity and other theories of gravity}",
      journal        = "Phys.Rev.",
      volume         = "D61",
      pages          = "084027",
      doi            = "10.1103/PhysRevD.61.084027",
      year           = "2000",
      eprint         = "gr-qc/9911095",
      archivePrefix  = "arXiv",
      primaryClass   = "gr-qc",
      SLACcitation   = "%%CITATION = GR-QC/9911095;%%",
}

@article{Compere:2013gja,
      author         = "Comp\`ere, Geoffrey and Song, Wei",
      title          = "{$\mathcal{W}$ symmetry and integrability of higher spin
                        black holes}",
      journal        = "JHEP",
      volume         = "1309",
      pages          = "144",
      doi            = "10.1007/JHEP09(2013)144",
      year           = "2013",
      eprint         = "1306.0014",
      archivePrefix  = "arXiv",
      primaryClass   = "hep-th",
      SLACcitation   = "%%CITATION = ARXIV:1306.0014;%%",
}

@article{Penington:2019npb,
    author = "Penington, Geoffrey",
    title = "{Entanglement Wedge Reconstruction and the Information Paradox}",
    eprint = "1905.08255",
    archivePrefix = "arXiv",
    primaryClass = "hep-th",
    doi = "10.1007/JHEP09(2020)002",
    journal = "JHEP",
    volume = "09",
    pages = "002",
    year = "2020"
}

@article{Penington:2019kki,
    author = "Penington, Geoff and Shenker, Stephen H. and Stanford, Douglas and Yang, Zhenbin",
    title = "{Replica wormholes and the black hole interior}",
    eprint = "1911.11977",
    archivePrefix = "arXiv",
    primaryClass = "hep-th",
    doi = "10.1007/JHEP03(2022)205",
    journal = "JHEP",
    volume = "03",
    pages = "205",
    year = "2022"
}

@article{Almheiri:2019psf,
    author = "Almheiri, Ahmed and Engelhardt, Netta and Marolf, Donald and Maxfield, Henry",
    title = "{The entropy of bulk quantum fields and the entanglement wedge of an evaporating black hole}",
    eprint = "1905.08762",
    archivePrefix = "arXiv",
    primaryClass = "hep-th",
    doi = "10.1007/JHEP12(2019)063",
    journal = "JHEP",
    volume = "12",
    pages = "063",
    year = "2019"
}

@article{Almheiri:2019qdq,
    author = "Almheiri, Ahmed and Hartman, Thomas and Maldacena, Juan and Shaghoulian, Edgar and Tajdini, Amirhossein",
    title = "{Replica Wormholes and the Entropy of Hawking Radiation}",
    eprint = "1911.12333",
    archivePrefix = "arXiv",
    primaryClass = "hep-th",
    doi = "10.1007/JHEP05(2020)013",
    journal = "JHEP",
    volume = "05",
    pages = "013",
    year = "2020"
}

@article{Bagchi:2013lma,
      author         = "Bagchi, Arjun and Detournay, Stephane and Grumiller,
                        Daniel and Simon, Joan",
      title          = "{Cosmic Evolution from Phase Transition of
                        Three-Dimensional Flat Space}",
      journal        = "Phys.Rev.Lett.",
      volume         = "111",
      pages          = "181301",
      doi            = "10.1103/PhysRevLett.111.181301",
      year           = "2013",
      eprint         = "1305.2919",
      archivePrefix  = "arXiv",
      primaryClass   = "hep-th",
      reportNumber   = "TUW-13-06",
      SLACcitation   = "%%CITATION = ARXIV:1305.2919;%%",
}

@article{Bagchi:2010zz,
      author         = "Bagchi, Arjun",
      title          = "{Correspondence between Asymptotically Flat Spacetimes
                        and Nonrelativistic Conformal Field Theories}",
      journal        = "Phys.Rev.Lett.",
      volume         = "105",
      pages          = "171601",
      doi            = "10.1103/PhysRevLett.105.171601",
      year           = "2010",
      SLACcitation   = "%%CITATION = PRLTA,105,171601;%%",
}

@article{Erices:2019onl,
    author = "Erices, Cristi{\'a}n and Riquelme, Miguel and Rodr{\'\i}guez, Pablo",
    title = "{BTZ black hole with Korteweg{\textendash}de Vries-type boundary conditions: Thermodynamics revisited}",
    eprint = "1907.13026",
    archivePrefix = "arXiv",
    primaryClass = "hep-th",
    doi = "10.1103/PhysRevD.100.126026",
    journal = "Phys. Rev. D",
    volume = "100",
    number = "12",
    pages = "126026",
    year = "2019"
}

@article{Kraus:2018xrn,
    author = "Kraus, Per and Liu, Junyu and Marolf, Donald",
    title = "{Cutoff AdS$_{3}$ versus the $ T\overline{T} $ deformation}",
    eprint = "1801.02714",
    archivePrefix = "arXiv",
    primaryClass = "hep-th",
    reportNumber = "CALT-TH-2018-002",
    doi = "10.1007/JHEP07(2018)027",
    journal = "JHEP",
    volume = "07",
    pages = "027",
    year = "2018"
}

@article{Apolo:2018qpq,
    author = "Apolo, Luis and Song, Wei",
    title = "{Strings on warped AdS$_{3}$ via $ \mathrm{T}\bar{\mathrm{J}} $ deformations}",
    eprint = "1806.10127",
    archivePrefix = "arXiv",
    primaryClass = "hep-th",
    doi = "10.1007/JHEP10(2018)165",
    journal = "JHEP",
    volume = "10",
    pages = "165",
    year = "2018"
}

@article{Chakraborty:2018vja,
    author = "Chakraborty, Soumangsu and Giveon, Amit and Kutasov, David",
    title = "{$ J\overline{T} $ deformed CFT$_{2}$ and string theory}",
    eprint = "1806.09667",
    archivePrefix = "arXiv",
    primaryClass = "hep-th",
    doi = "10.1007/JHEP10(2018)057",
    journal = "JHEP",
    volume = "10",
    pages = "057",
    year = "2018"
}

@article{Faulkner:2010jy,
    author = "Faulkner, Thomas and Liu, Hong and Rangamani, Mukund",
    title = "{Integrating out geometry: Holographic Wilsonian RG and the membrane paradigm}",
    eprint = "1010.4036",
    archivePrefix = "arXiv",
    primaryClass = "hep-th",
    reportNumber = "MIT-CTP-4185, DCPT-10-47",
    doi = "10.1007/JHEP08(2011)051",
    journal = "JHEP",
    volume = "08",
    pages = "051",
    year = "2011"
}

@article{Heemskerk:2010hk,
    author = "Heemskerk, Idse and Polchinski, Joseph",
    title = "{Holographic and Wilsonian Renormalization Groups}",
    eprint = "1010.1264",
    archivePrefix = "arXiv",
    primaryClass = "hep-th",
    doi = "10.1007/JHEP06(2011)031",
    journal = "JHEP",
    volume = "06",
    pages = "031",
    year = "2011"
}

@article{Adami:2025pqr,
    author = "Adami, H. and Sheikh-Jabbari, M. M. and Taghiloo, V.",
    title = "{Gravity Is Induced By Renormalization Group Flow}",
    eprint = "2508.09633",
    archivePrefix = "arXiv",
    primaryClass = "hep-th",
    month = "8",
    year = "2025"
}

@article{Yu:2022bcp,
    author = "Yu, Zhe-fei and Chen, Bin",
    title = "{Free field realization of the BMS Ising model}",
    eprint = "2211.06926",
    archivePrefix = "arXiv",
    primaryClass = "hep-th",
    doi = "10.1007/JHEP08(2023)116",
    journal = "JHEP",
    volume = "08",
    pages = "116",
    year = "2023"
}

@article{Guica:2017lia,
    author = "Guica, Monica",
    title = "{An integrable Lorentz-breaking deformation of two-dimensional CFTs}",
    eprint = "1710.08415",
    archivePrefix = "arXiv",
    primaryClass = "hep-th",
    doi = "10.21468/SciPostPhys.5.5.048",
    journal = "SciPost Phys.",
    volume = "5",
    number = "5",
    pages = "048",
    year = "2018"
}

@article{Barnich:2006av,
      author         = "Barnich, Glenn and Compere, Geoffrey",
      title          = "{Classical central extension for asymptotic symmetries at
                        null infinity in three spacetime dimensions}",
      journal        = "Class.Quant.Grav.",
      volume         = "24",
      pages          = "F15-F23",
      doi            = "10.1088/0264-9381/24/5/F01, 10.1088/0264-9381/24/11/C01",
      year           = "2007",
      eprint         = "gr-qc/0610130",
      archivePrefix  = "arXiv",
      primaryClass   = "gr-qc",
      reportNumber   = "ULB-TH-06-08",
      SLACcitation   = "%%CITATION = GR-QC/0610130;%%",
}

@article{Bagchi:2012yk,
      author         = "Bagchi, Arjun and Detournay, Stephane and Grumiller,
                        Daniel",
      title          = "{Flat-Space Chiral Gravity}",
      journal        = "Phys.Rev.Lett.",
      volume         = "109",
      pages          = "151301",
      doi            = "10.1103/PhysRevLett.109.151301",
      year           = "2012",
      eprint         = "1208.1658",
      archivePrefix  = "arXiv",
      primaryClass   = "hep-th",
      SLACcitation   = "%%CITATION = ARXIV:1208.1658;%%",
}

@Article{comment,
 author = {Rador, T. and Arapoglu, S. and Semiz, I.},
 title = {Comment on ``Model for Gravity at Large Distances''},
 year = {2012},
}

@article{Campoleoni:2011hg,
      author         = "Campoleoni, Andrea and Fredenhagen, Stefan and
                        Pfenninger, Stefan",
      title          = "{Asymptotic W-symmetries in three-dimensional higher-spin
                        gauge theories}",
      journal        = "JHEP",
      volume         = "1109",
      pages          = "113",
      year           = "2011",
      eprint         = "1107.0290",
      archivePrefix  = "arXiv",
      primaryClass   = "hep-th",
      reportNumber   = "AEI-2011-041",
      SLACcitation   = "%%CITATION = ARXIV:1107.0290;%%"
}

@article{Coussaert:1995zp,
      author         = "Coussaert, Oliver and Henneaux, Marc and van Driel,
                        Peter",
      title          = "{The Asymptotic dynamics of three-dimensional Einstein
                        gravity with a negative cosmological constant}",
      journal        = "Class.Quant.Grav.",
      volume         = "12",
      pages          = "2961-2966",
      doi            = "10.1088/0264-9381/12/12/012",
      year           = "1995",
      eprint         = "gr-qc/9506019",
      archivePrefix  = "arXiv",
      primaryClass   = "gr-qc",
      reportNumber   = "ULB-TH-95-08",
      SLACcitation   = "%%CITATION = GR-QC/9506019;%%"
}

@book{marchenko2013sturm,
  title={Sturm-Liouville operators and applications},
  author={Marchenko, Vladimir Aleksandrovich},
  year={2013},
  publisher={Springer-Verlag}
}

@book{dickey2003soliton,
  title={Soliton equations and Hamiltonian systems},
  author={Dickey, Leonid A},
  volume={26},
  year={2003},
  publisher={World scientific}
}

@article{lifshitz1941theory,
  title={On the theory of second-order phase transitions I \& II},
  author={Lifshitz, EM},
  journal={Zh. Eksp. Teor. Fiz},
  volume={11},
  number={255},
  pages={269},
  year={1941}
}

@book{Dickey:1991xa,
    author = "Dickey, L. A.",
    title = "{Soliton equations and Hamiltonian systems}",
    volume = "12",
    year = "1991"
}

@article{Gelfand:1975rn,
    author = "Gelfand, I. M. and Dikii, L. A.",
    title = "{Asymptotic behavior of the resolvent of Sturm-Liouville equations and the algebra of the Korteweg-De Vries equations}",
    doi = "10.1070/RM1975v030n05ABEH001522",
    journal = "Russ. Math. Surveys",
    volume = "30",
    number = "5",
    pages = "77--113",
    year = "1975"
}

@article{magri1978simple,
  title={A simple model of the integrable Hamiltonian equation},
  author={Magri, Franco},
  journal={Journal of Mathematical Physics},
  volume={19},
  number={5},
  pages={1156--1162},
  year={1978},
  publisher={American Institute of Physics}
}

@article{Campoleoni:2010zq,
      author         = "Campoleoni, Andrea and Fredenhagen, Stefan and
                        Pfenninger, Stefan and Theisen, Stefan",
      title          = "{Asymptotic symmetries of three-dimensional gravity
                        coupled to higher-spin fields}",
      journal        = "JHEP",
      volume         = "1011",
      pages          = "007",
      doi            = "10.1007/JHEP11(2010)007",
      year           = "2010",
      eprint         = "1008.4744",
      archivePrefix  = "arXiv",
      primaryClass   = "hep-th"
}

@Article{Achucarro:1987vz,
     author    = "Achucarro, A. and Townsend, P. K.",
     title     = "A {C}HERN-{S}IMONS ACTION FOR THREE-DIMENSIONAL {A}NTI-DE {S}ITTER
                  SUPERGRAVITY THEORIES",
     journal   = "Phys. Lett.",
     volume    = "B180",
     year      = 1986,
     pages     = 89,
     doi = "10.1016/0370-2693(86)90140-1",
     SLACcitation  = "%%CITATION = PHLTA,B180,89;%%"
}

@Article{Balasubramanian:1999re,
     author    = "Balasubramanian, Vijay and Kraus, Per",
     title     = "A stress tensor for anti-de {S}itter gravity",
     journal   = "Commun. Math. Phys.",
     volume    = "208",
     year      = "1999",
     pages     = "413-428",
     eprint    = "hep-th/9902121",
     SLACcitation  = "%%CITATION = HEP-TH 9902121;%%"
}

@Article{Banados:1994tn,
     author    = "Ba\~nados, Maximo",
     title     = "{Global charges in Chern-Simons field theory and the (2+1)
                  black hole}",
     journal   = "Phys. Rev.",
     volume    = "D52",
     year      = "1995",
     pages     = "5816",
     eprint    = "hep-th/9405171",
     SLACcitation  = "%%CITATION = HEP-TH 9405171;%%"
}

@Article{Banados:1998gg,
     author    = "Ba\~nados, Maximo",
     title     = "{Three-dimensional quantum geometry and black holes}",
     year      = "1998",
     eprint    = "hep-th/9901148",
     archivePrefix = "arXiv",
     SLACcitation  = "%%CITATION = HEP-TH/9901148;%%"
}

@Article{Barcelo:2005fc,
     author    = "Barcelo, Carlos and Liberati, Stefano and Visser, Matt",
     title     = "Analogue gravity",
     journal   = "Living Rev. Rel.",
     volume    = "8",
     year      = "2005",
     pages     = "12",
     eprint    = "gr-qc/0505065",
     SLACcitation  = "%%CITATION = GR-QC 0505065;%%"
}

@Article{Bianchi:2001kw,
     author    = "Bianchi, Massimo and Freedman, Daniel Z. and Skenderis,
                  Kostas",
     title     = "{Holographic Renormalization}",
     journal   = "Nucl. Phys.",
     volume    = "B631",
     year      = "2002",
     pages     = "159-194",
     eprint    = "hep-th/0112119",
     archivePrefix = "arXiv",
     SLACcitation  = "%%CITATION = HEP-TH/0112119;%%"
}

@Article{Brown:1986nw,
     author    = "Brown, J. David and Henneaux, M.",
     title     = "{Central Charges in the Canonical Realization of Asymptotic
                  Symmetries: An Example from Three-Dimensional Gravity}",
     journal   = "Commun. Math. Phys.",
     volume    = "104",
     year      = "1986",
     pages     = "207-226",
     SLACcitation  = "%%CITATION = CMPHA,104,207;%%"
}

@Article{Brown:1992br,
     author    = "Brown, J. David and York, Jr., James W.",
     title     = "Quasilocal energy and conserved charges derived from the
                  gravitational action",
     journal   = "Phys. Rev.",
     volume    = "D47",
     year      = "1993",
     pages     = "1407-1419",
     SLACcitation  = "%%CITATION = PHRVA,D47,1407;%%"
}

@article{Adami:2025pfk,
    author = "Adami, Hamed and Latifi, Anouchah",
    title = "{Integrability in Three-Dimensional Gravity: Eigenfunction-Forced KdV Flows}",
    eprint = "2510.10519",
    archivePrefix = "arXiv",
    primaryClass = "hep-th",
    month = "10",
    year = "2025"
}

@article{Ojeda:2020bgz,
    author = "Ojeda, Emilio and P{\'e}rez, Alfredo",
    title = "{Integrable systems and the boundary dynamics of higher spin gravity on AdS$_{3}$}",
    eprint = "2009.07829",
    archivePrefix = "arXiv",
    primaryClass = "hep-th",
    reportNumber = "CECS-PHY-20/02",
    doi = "10.1007/JHEP11(2020)089",
    journal = "JHEP",
    volume = "11",
    pages = "089",
    year = "2020"
}

@Article{Carlip:2005zn,
     author    = "Carlip, S.",
     title     = "{Conformal field theory, (2+1)-dimensional gravity, and the
                  BTZ black hole}",
     journal   = "Class. Quant. Grav.",
     volume    = "22",
     year      = "2005",
     pages     = "R85-R124",
     eprint    = "gr-qc/0503022",
     SLACcitation  = "%%CITATION = GR-QC 0503022;%%"
}

@article{Compere:2018aar,
      author         = "Compère, Geoffrey and Fiorucci, Adrien",
      title          = "{Advanced Lectures on General Relativity}",
      journal        = "Lect. Notes Phys.",
      volume         = "952",
      year           = "2019",
      pages          = "150",
      doi            = "10.1007/978-3-030-04260-8",
      eprint         = "1801.07064",
      archivePrefix  = "arXiv",
      primaryClass   = "hep-th",
      SLACcitation   = "%%CITATION = ARXIV:1801.07064;%%"
}

@Article{Gubser:1998bc,
     author    = "Gubser, S. S. and Klebanov, Igor R. and Polyakov, Alexander
                  M.",
     title     = "Gauge theory correlators from non-critical string theory",
     journal   = "Phys. Lett.",
     volume    = "B428",
     year      = "1998",
     pages     = "105-114",
     eprint    = "hep-th/9802109",
     SLACcitation  = "%%CITATION = HEP-TH 9802109;%%"
}

@Article{Iyer:1994ys,
     author    = "Iyer, Vivek and Wald, Robert M.",
     title     = "Some properties of {N}{\"o}ther charge and a proposal for
                  dynamical black hole entropy",
     journal   = "Phys. Rev.",
     volume    = "D50",
     year      = "1994",
     pages     = "846-864",
     archive   = "http://arXiv.org/abs",
     eprint    = "gr-qc/9403028",
     SLACcitation  = "%%CITATION = GR-QC 9403028;%%"
}

@article{Lee:1990nz,
      author         = "Lee, J. and Wald, Robert M.",
      title          = "{Local symmetries and constraints}",
      journal        = "J. Math. Phys.",
      volume         = "31",
      year           = "1990",
      pages          = "725-743",
      doi            = "10.1063/1.528801",
      SLACcitation   = "%%CITATION = JMAPA,31,725;%%"
}

@Article{Maldacena:1997re,
     author    = "Maldacena, Juan M.",
     title     = "{The large $N$ limit of superconformal field theories and
                  supergravity}",
     journal   = "Adv. Theor. Math. Phys.",
     volume    = "2",
     year      = "1998",
     pages     = "231-252",
     eprint    = "hep-th/9711200",
     SLACcitation  = "%%CITATION = HEP-TH 9711200;%%"
}

@Article{Regge:1974zd,
     author    = "Regge, Tullio and Teitelboim, Claudio",
     title     = "ROLE OF SURFACE INTEGRALS IN THE {H}AMILTONIAN FORMULATION OF
                  GENERAL RELATIVITY",
     journal   = "Ann. Phys.",
     volume    = "88",
     year      = "1974",
     pages     = "286",
     SLACcitation  = "%%CITATION = APNYA,88,286;%%"
}

@Article{Skenderis:2002wp,
     author    = "Skenderis, Kostas",
     title     = "Lecture notes on holographic renormalization",
     journal   = "Class. Quant. Grav.",
     volume    = "19",
     year      = "2002",
     pages     = "5849-5876",
     eprint    = "hep-th/0209067",
     SLACcitation  = "%%CITATION = HEP-TH 0209067;%%"
}

@Article{Strominger:1997eq,
     author    = "Strominger, Andrew",
     title     = "Black hole entropy from near-horizon microstates",
     journal   = "JHEP",
     volume    = "02",
     year      = "1998",
     pages     = "009",
     eprint    = "hep-th/9712251",
     SLACcitation  = "%%CITATION = HEP-TH 9712251;%%"
}

@article{Grumiller:2017sjh,
    author = "Grumiller, Daniel and Merbis, Wout and Riegler, Max",
    title = "{Most general flat space boundary conditions in three-dimensional Einstein gravity}",
    eprint = "1704.07419",
    archivePrefix = "arXiv",
    primaryClass = "hep-th",
    reportNumber = "TUW-17-04",
    doi = "10.1088/1361-6382/aa8004",
    journal = "Class. Quant. Grav.",
    volume = "34",
    number = "18",
    pages = "184001",
    year = "2017"
}

@Article{Witten:1988hc,
     author    = "Witten, Edward",
     title     = "(2+1)-DIMENSIONAL GRAVITY AS AN EXACTLY SOLUBLE SYSTEM",
     journal   = "Nucl. Phys.",
     volume    = "B311",
     year      = "1988",
     pages     = "46",
     doi = "10.1016/0550-3213(88)90143-5",
     SLACcitation  = "%%CITATION = NUPHA,B311,46;%%"
}

@Article{Witten:1998qj,
     author    = "Witten, Edward",
     title     = "{Anti-de Sitter space and holography}",
     journal   = "Adv. Theor. Math. Phys.",
     volume    = "2",
     year      = "1998",
     pages     = "253-291",
     eprint    = "hep-th/9802150",
     SLACcitation  = "%%CITATION = HEP-TH 9802150;%%"
}

@Article{deBoer:1999xf,
     author    = "de Boer, Jan and Verlinde, Erik P. and Verlinde, Herman L.",
     title     = "On the holographic renormalization group",
     journal   = "JHEP",
     volume    = "08",
     year      = "2000",
     pages     = "003",
     eprint    = "hep-th/9912012",
     SLACcitation  = "%%CITATION = HEP-TH 9912012;%%"
}

@article{Grumiller:2016pqb,
    author = "Grumiller, Daniel and Riegler, Max",
    title = "{Most general AdS$_{3}$ boundary conditions}",
    eprint = "1608.01308",
    archivePrefix = "arXiv",
    primaryClass = "hep-th",
    reportNumber = "TUW-16-16",
    doi = "10.1007/JHEP10(2016)023",
    journal = "JHEP",
    volume = "10",
    pages = "023",
    year = "2016"
}

@article{Hartman:2018tkw,
    author = "Hartman, Thomas and Kruthoff, Jorrit and Shaghoulian, Edgar and Tajdini, Amirhossein",
    title = "{Holography at finite cutoff with a $T^2$ deformation}",
    eprint = "1807.11401",
    archivePrefix = "arXiv",
    primaryClass = "hep-th",
    doi = "10.1007/JHEP03(2019)004",
    journal = "JHEP",
    volume = "03",
    pages = "004",
    year = "2019"
}

@article{Taylor:2018xcy,
    author = "Taylor, Marika",
    title = "{$T \bar{T}$ deformations in general dimensions}",
    eprint = "1805.10287",
    archivePrefix = "arXiv",
    primaryClass = "hep-th",
    doi = "10.4310/ATMP.2023.v27.n1.a2",
    journal = "Adv. Theor. Math. Phys.",
    volume = "27",
    number = "1",
    pages = "37--63",
    year = "2023"
}

@article{McGough:2016lol,
    author = "McGough, Lauren and Mezei, M\'ark and Verlinde, Herman",
    title = "{Moving the CFT into the bulk with $ T\overline{T} $}",
    eprint = "1611.03470",
    archivePrefix = "arXiv",
    primaryClass = "hep-th",
    doi = "10.1007/JHEP04(2018)010",
    journal = "JHEP",
    volume = "04",
    pages = "010",
    year = "2018"
}

@article{Guica:2019nzm,
    author = "Guica, Monica and Monten, Ruben",
    title = "{$T\bar T$ and the mirage of a bulk cutoff}",
    eprint = "1906.11251",
    archivePrefix = "arXiv",
    primaryClass = "hep-th",
    doi = "10.21468/SciPostPhys.10.2.024",
    journal = "SciPost Phys.",
    volume = "10",
    number = "2",
    pages = "024",
    year = "2021"
}

@article{Cavaglia:2016oda,
    author = "Cavagli\`a, Andrea and Negro, Stefano and Sz\'ecs\'enyi, Istv\'an M. and Tateo, Roberto",
    title = "{$T \bar{T}$-deformed 2D Quantum Field Theories}",
    eprint = "1608.05534",
    archivePrefix = "arXiv",
    primaryClass = "hep-th",
    doi = "10.1007/JHEP10(2016)112",
    journal = "JHEP",
    volume = "10",
    pages = "112",
    year = "2016"
}

@article{Aharony:2018bad,
    author = "Aharony, Ofer and Datta, Shouvik and Giveon, Amit and Jiang, Yunfeng and Kutasov, David",
    title = "{Modular invariance and uniqueness of $T\bar{T}$ deformed CFT}",
    eprint = "1808.02492",
    archivePrefix = "arXiv",
    primaryClass = "hep-th",
    doi = "10.1007/JHEP01(2019)086",
    journal = "JHEP",
    volume = "01",
    pages = "086",
    year = "2019"
}

@article{Latifi:2026ttmp,
  author  = {Latifi, Anouchah},
  title   = {{Geometric Leakage versus Dissipation: Open Boundaries in Chern--Simons Fluid Dynamics}},
  journal = {Trans. Theor. Math. Phys.},
  year    = {2026},
  pages   = {82--89},
  issn    = {3041-8984},
  doi     = {10.30511/ttmp.2026.2091721.1081}
}

@article{Chen:2025jzb,
    author = "Chen, Liangyu and Du, Zhengyuan and Liu, Kangning and Song, Wei",
    title = "{Symmetries and operators in $T\bar{T}$ deformed CFTs}",
    eprint = "2507.08588",
    archivePrefix = "arXiv",
    primaryClass = "hep-th",
    month = "7",
    year = "2025"
}

@article{Cardenas:2025qqi,
    author = "C{\'a}rdenas, Marcela and Correa, Francisco and Pino, Miguel",
    title = "{Integrable black hole dynamics in the asymptotic structure of AdS$_{3}$}",
    eprint = "2504.20292",
    archivePrefix = "arXiv",
    primaryClass = "hep-th",
    month = "4",
    year = "2025"
}

@article{Cardenas:2024hah,
	author = "C{\'a}rdenas, Marcela",
	title = "{KdV conformal symmetry breaking in nearly AdS$_{2}$}",
	eprint = "2405.03128",
	archivePrefix = "arXiv",
	primaryClass = "hep-th",
	doi = "10.1007/JHEP10(2024)052",
	journal = "JHEP",
	volume = "10",
	pages = "052",
	year = "2024"
}

@article{Pino:2025crn,
	author = "Pino, Miguel and Reyes, Francisco",
	title = "{Non-axisymmetric (2+1) black holes with Dym boundary conditions}",
	eprint = "2511.06567",
	archivePrefix = "arXiv",
	primaryClass = "hep-th",
	month = "11",
	year = "2025"
}

@article{Castro:2025itb,
    author = "Castro, Alejandra and Mancilla, Robinson and Papadimitriou, Ioannis",
    title = "{Near-extremal dynamics away from the horizon}",
    eprint = "2507.01126",
    archivePrefix = "arXiv",
    primaryClass = "hep-th",
    doi = "10.1007/JHEP11(2025)083",
    journal = "JHEP",
    volume = "11",
    pages = "083",
    year = "2025"
}

@article{Melnikov:2018fhb,
    author = "Melnikov, Dmitry and Novaes, F{\'a}bio and P{\'e}rez, Alfredo and Troncoso, Ricardo",
    title = "{Lifshitz Scaling, Microstate Counting from Number Theory and Black Hole Entropy}",
    eprint = "1808.04034",
    archivePrefix = "arXiv",
    primaryClass = "hep-th",
    reportNumber = "CECS-PHY-18/02, ITEP-TH-20/18",
    doi = "10.1007/JHEP06(2019)054",
    journal = "JHEP",
    volume = "06",
    pages = "054",
    year = "2019"
}


\end{document}